\documentclass[aps,pre,twocolumn,showpacs]{revtex4}

\usepackage{graphicx}
\usepackage{amsmath,amscd,amssymb}

\newcommand{\be}{\begin{equation}}
\newcommand{\ee}{\end{equation}}
\newcommand{\bea}{\begin{eqnarray}}
\newcommand{\eea}{\end{eqnarray}}
\newcommand{\br}{\mathbf{r}}
\newcommand{\bk}{\mathbf{k}}

\begin{document}

\title{Variational approach for electrolyte solutions:\\ from dielectric interfaces to charged nanopores}

\author{Sahin Buyukdagli\footnote{email:~\texttt{buyuk@irsamc.ups-tlse.fr}}, Manoel Manghi\footnote{email:~\texttt{manghi@irsamc.ups-tlse.fr}}, and John Palmeri\footnote{email:~\texttt{john.palmeri@irsamc.ups-tlse.fr}}}
\affiliation{Universit\'e de Toulouse; UPS; \\ Laboratoire de Physique Th\'eorique (IRSAMC); F-31062 Toulouse, France}
\affiliation{CNRS; LPT (IRSAMC); F-31062 Toulouse, France}
\date{\today}

\begin{abstract}
A variational theory is developed to study electrolyte solutions,
composed of interacting point-like ions in a solvent, in the
presence of dielectric discontinuities and charges at the
boundaries. Three important and non-linear electrostatic effects
induced by these interfaces are taken into account: surface charge
induced electrostatic field, solvation energies due to the ionic
cloud, and image charge repulsion. Our variational equations thus go
beyond the mean-field theory, or weak coupling limit, where thermal
fluctuations overcome electrostatic correlations, and allows one to
reach the opposite strong coupling limit, where electrostatic
interactions induced by interfaces dominate. The influence of salt
concentration, ion valency, dielectric jumps, and surface charge is
studied in two geometries. i) A single neutral dielectric interface
(e.g. air--water or electrolyte--membrane) with an asymmetric
electrolyte. A charge separation and thus an electrostatic field
gets established due to the different image charge repulsions for
coions and counterions. Both charge distributions and surface
tension are computed and compared to previous approximate
calculations. For symmetric electrolyte solutions close to a charged
surface, two zones are characterized. In the first one, in contact
with the surface and with size proportional to the logarithm of the
coupling parameter, strong image forces and strong coupling impose a
total ion exclusion, while in the second zone the mean-field
approach applies. ii) A symmetric electrolyte confined between two
dielectric interfaces as a simple model of ion rejection from
nanopores in membranes. The competition between image charge
repulsion and attraction of counterions by the membrane charge is
studied. For small surface charge, the counterion partition
coefficient decreases with increasing pore size up to a critical
pore size, contrary to neutral membranes. For larger pore sizes, the
whole system behaves like a neutral pore. For strong coupling and
small pore size, coion exclusion is total and the counterion
partition coefficient is solely determined by global
electroneutrality. A quantitative comparison is made with a previous
approach, where image and surface charge effects were smeared out in
the pore. It is shown that the variational method allows one to go
beyond the constant Donnan potential approximation, with deviations
stronger at high ion concentrations or small pore sizes. The
prediction of the variational method is also compared with MC
simulations and a good agreement is observed.

\end{abstract}

\maketitle
\section{Introduction}

The first experimental evidence for the enhancement of the surface
tension of inorganic salt solutions compared to that of pure water
was obtained more than eight decades ago~\cite{heyd,weiss}. Wagner
proposed the correct physical picture~\cite{wagner} by relating this
effect to image forces that originate from the dielectric
discontinuity and act on ions close to the water--air interface. He
also correctly pointed out the fundamental importance of the ionic
screening of image forces and formulated a theoretical description
of the problem by establishing a differential equation for the
electrostatic potential and solving it numerically to compute the
surface tension. Using series expansions, Onsager and Samaras found
the celebrated limiting law~\cite{onsager} that relates the surface
tension of symmetric electrolytes to the bulk electrolyte density at
low salt concentration. However, it is known that the consideration
of charge asymmetry leads to a technical complication. Indeed, image
charge repulsion, whose amplitude is proportional to the square of
ion valency, leads to a split of concentration profiles for ions of
different charge, which in turn causes a local violation of the
electroneutrality and induces an electrostatic field close to a
neutral dielectric interface. Bravina derived five decades ago a
Poisson-Boltzmann type of equation for this field~\cite{bravina} and
used several approximations in order to derive integral expressions
for the charge distribution and the surface tension.

These image charge forces play also a key role in slit-like
nanopores which are model systems for studying ion rejection and
nanofiltration by porous membranes (see the review~\cite{yarosch}
and references therein, and ~\cite{Schoch} for a review of
nano-fluidics). Several results have been found in this geometry and
also for cylindrical nanopores beyond the mean-field approach (using
the Debye closure and the BBGKY hierarchical equations) and
averaging all dielectric and charge effects over the pore cross
section. Within these two approximations, the salt reflection
coefficient has been studied as a function of the pore size, the
bulk salt concentration and the pore surface charge.

More precisely, the strength of electrostatic correlations of ions
in the presence of charged interfaces \textit{without dielectric
discontinuity} is quantified by one unique coupling parameter \be
\Xi=2\pi q^3 \ell_B^2\sigma_s \ee where $q$ is the ion valency, and
$\sigma_s$ the fixed surface charge~\cite{MorI,MorII,hoda}. The
Bjerrum length in water for monovalent ions,
$\ell_B=e^2/(4\pi\epsilon_w k_BT)\approx 0.7$~nm ($\epsilon_w$ is
the dielectric permittivity of water) is defined as the distance at
which the electrostatic interaction between two elementary charges
is equal to the thermal energy $k_BT$. The second characteristic
length is the Gouy-Chapman length $\ell_G=1/(2\pi q \ell_B\sigma_s)$
defined as the distance at which the electrostatic interaction
between a single ion and a charged interface is equal to $k_BT$. The
coupling parameter can be reexpressed in terms of these two lengths
as $\Xi=q^2\ell_B/\ell_G$. On the one hand, the limit $\Xi\to 0$,
called the weak coupling (WC) limit, is where the physics of the
Coulomb system is governed by the mean-field or Poisson-Boltzmann
(PB) theory, and  thermal fluctuations overcome electrostatic
interactions. It describes systems characterized by a high
temperature, low ion valency or weak surface charge. On the other
hand, $\Xi\to \infty$ is the strong coupling (SC) limit,
corresponding to low temperature, high valency of mobile ions or
strong surface charge. In this limit, ion--charged surface
interactions control the ion distribution perpendicularly to the
interface. For single interface and slab geometries, several
perturbative approaches going beyond the WC
limit~\cite{netzorland_PB} or below the SC
limit~\cite{MorI,moreira_MC,Review} have been developed. Although
these calculations were able to capture important phenomena such as
charge renormalization~\cite{Alex}, ion specific effects at the
water-air interface~\cite{Lev1,Lev2}, Manning
condensation~\cite{Man}, effect of monopoles~\cite{KanducDip} or
attraction between similarly charged objects, they also showed slow
convergence properties, which indicates the inability of high-order
expansions to explore the intermediate regime, $\Xi\simeq1$. This is
quite frustrating since the common experimental situation usually
corresponds to the range $0.1<\Xi<10$ where neither WC nor SC theory
is totally adequate.

Consequently, a non-perturbative approach valid for the whole range
of $\Xi$ is needed. A first important attempt in this direction has
been made by Netz and Orland~\cite{netz_var} who derived variational
equations within the primitive model for point-like ions and solved
them at the mean-field level in order to illustrate charge
renormalization effect. Interestingly, these differential equations
are equivalent to the closure equations established in the context
of electrolytes in nanopores~\cite{yarosch}. They are too
complicated to be solved analytically or even numerically for
general $\Xi$. A few years later, Curtis and Lue~\cite{curtis} and
Hatlo \textit{et al.}~\cite{hatlo} investigated the partition of
symmetric electrolytes at neutral dielectric surfaces using a
similar variational approach (see also the
review~\cite{hatlo_review}). They have also recently proposed a new
variational scheme based on a hybrid low fugacity and mean-field
expansion~\cite{hatlo2}, and showed that their approach agrees well
with Monte-Carlo simulation results for the counterions-only case.
However, this method is quite difficult to handle, and one has to
solve two coupled variational equations, i.e. a sixth order
differential equation for the external potential together with a
second algebraic equation.  Within this approach, these authors
generalized the study of ion-ion correlations for counterions close
to a charged dielectric interface, first done by Netz in the WC and
SC limits~\cite{netz_WSC}, to intermediate values of $\Xi$. They
also studied an electrolyte between two charged surfaces without
dielectric discontinuities at the pore boundary, in two cases:
counterions only and added salt, handled at the mean-field
level~\cite{Arx}. Although this simplification allows one to focus
exclusively on ion-ion correlations induced by the surface charge,
the dielectric discontinuity can not be discarded in synthetic or
biological membranes. Indeed, it is known that image forces play a
crucial in ion filtration mechanisms~\cite{yarosch}. The main goal
of this work is to propose a variational analysis which is simple
enough to intuitively illustrate ionic exclusion in slit pores, by
focusing on the competition between image charge repulsion and
surface charge interaction. Moreover, our approach allows us to
connect nanofiltration studies~\cite{starov,lefebvreII,YarII} with
field-theoretic approaches of confined electrolyte solutions within
a generalized Onsager-Samaras approximation~\cite{onsager}
characterized by a uniform variational screening length. This
variational parameter takes into account the interaction with both
image charge and surface charge. We also compare the prediction of
the variational theory with Monte-Carlo simulations~\cite{Li} and
show that the agreement is good.

The paper is organized as follows. The variational formalism for
Coulombic systems in the presence of dielectric discontinuities is
introduced in Section~\ref{varcalc}. Section~\ref{singleinter} deals
with a single interface. We show that the introduction of simple
variational potentials allows one to fully account for the physics
of asymmetric electrolytes at dielectric interfaces (e.g.
water--air, liquid--liquid and liquid--solid interfaces, see
ref.~\cite{Benji}), first studied by Bravina~\cite{bravina} using
several approximations, as well as the case of charged surfaces. In
Section~\ref{dbleinter}, the variational approach is applied to a
symmetric electrolyte confined between two dielectric surfaces in
order to investigate the problem of ion rejection from membrane
nanopores. Using restricted variational potentials, we show that due
to the interplay between image charge repulsion and direct
electrostatic interaction with the charged surface, the ionic
partition coefficient has a non-monotonic behaviour as a function of
pore size.

\section{Variational Calculation}
\label{varcalc}

In this section, the field theoretic variational approach for many
body systems composed of point-like ions in the presence of
dielectric interfaces is presented. Since the field theoretic
formalism as well as the first order variational scheme have already
been introduced in previous works~\cite{netz_var,curtis}, we only
illustrate the general lines.

The grand-canonical partition function of $p$ ion species in a
liquid of spatially varying dielectric constant $\epsilon(\br)$ is
\be\label{ZG} \mathcal{Z}=\prod_{i=1}^p\sum_{N_i=
0}^\infty\frac{e^{N_i\mu_i}}{N_i!\lambda_t^{3N_i}}\int\prod_{j=1}^{N_i}\mathrm{d}\br_{ij}e^{-(H-E_s)}
\ee where $\lambda_t$ is the thermal wavelength of an ion, $\mu_i$
denotes the chemical potential and $N_i$ the total number of ions of
type $i$. For sake of simplicity, all energies are expressed in
units of $k_BT$. The electrostatic interaction is \be
H=\frac{1}{2}\int
\mathrm{d}\br'\mathrm{d}\br\rho_c(\br)v_c(\br,\br')\rho_c(\br') \ee
where $\rho_c$ is the charge distribution (in units of $e$) \be
\rho_c(\br)=\sum_{i=1}^p\sum_{j=1}^{N_i}q_i\delta(\br-\br_{ij})+\rho_s(\br),
\ee and $q_i$ denotes the valency of each species, $\rho_s(\br)$
stands for the fixed charge distribution and $v_c(\br,\br')$ is the
Coulomb potential whose inverse is defined as \be
v_c^{-1}(\br,\br')=-\frac{k_BT}{e^2}\nabla\left[\epsilon(\br)\nabla\delta(\br-\br')\right]
\label{coulomb} \ee where $\epsilon(\br)$ is a spatially varying
permittivity. The self-energy of mobile ions, which is subtracted
from the total electrostatic energy, is \be
E_s=\frac{v^b_c(0)}{2}\sum_{i=1}^pN_iq_i^2 \ee where
$v^b_c(\br)=\ell_B/r$ is the bare Coulomb potential  for
$\epsilon(\br)=\epsilon_w$. After performing a Hubbard-Stratonovitch
transformation and the summation over $N_i$ in Eq.~(\ref{ZG}), the
grand-canonical partition function takes the form of a functional
integral over an imaginary electrostatic auxiliary field
$\phi(\br)$, $\mathcal{Z}=\int \mathcal{D}\phi\;e^{-H[\phi]}$. The
Hamiltonian is \bea H[\phi]&=&\int
\mathrm{d}\br\left[\frac{\left[\nabla\phi(\br)\right]^2}{8\pi
\ell_B(\br)}-i\rho_s(\br)\phi(\br)\right.\\ &&\left.-
\sum_i\lambda_i e^{\frac{q_i^2}{2}v_c^b(0)+i q_i
\phi(\br)}\right]\nonumber \eea where a rescaled fugacity \be
\lambda_i=e^{\mu_i}/\lambda_t^3 \ee has been introduced. The
variational method consists in optimizing the first order cumulant
\be\label{Fvar} F_v=F_0+\langle H-H_0\rangle_0. \ee where averages
$\langle\cdots\rangle_0$ are to be evaluated with respect to the
most general Gaussian Hamiltonian~\cite{netz_var}, \be\label{HVar}
H_0[\phi]=\frac{1}{2}\int_{\br,\br'}\left[\phi(\br)-i\phi_0(\br)\right]v^{-1}_0(\br,\br')\left[\phi(\br')-i\phi_0(\br')\right]
\ee and $F_0=-\frac{1}{2}\mathrm{tr}\ln v_0$. The variational
principle consists in looking for the optimal choices of the
electrostatic kernel $v_0(\br,\br')$ and the average electrostatic
potential $\phi_0(\br)$ which extremize the variational grand
potential Eq.~(\ref{Fvar}). The variational equations $\delta
F_v/\delta v_0^{-1}(\br,\br')=0$ and $\delta F_v/\delta
\phi_0(\br)=0$, for a symmetric electrolyte and
$\epsilon(\br)=\epsilon_w$, yield
   \bea
\Delta \phi_0(\br)&-&8\pi \ell_Bq \lambda
e^{-\frac{q^2}{2}W(\br)}\sinh\left[q\phi_0(\br)\right]\nonumber\\
&&\hspace{2.5cm}=-4\pi \ell_B\rho_s(\br)\label{VarNetzP}\\
-\Delta v_0(\br,\br')&+&8\pi \ell_Bq^2\lambda  e^{-\frac{q^2}{2}
W(\br)}\cosh\left[q\phi_0(\br)\right] v_0(\br,\br')\nonumber\\
&&\hspace{2.5cm}=4\pi \ell_B\delta(\br-\br').\label{VarNetzG}
   \eea
where we have defined \be W(\br) \equiv \lim_{\br\to
\br'}\left[v_0(\br,\br')-v_c^b(\br-\br')\right] \ee whose physical
signification will be given below. The second terms on the LHS of
Eq.~(\ref{VarNetzP}) and of Eq.~(\ref{VarNetzG}) have simple
physical interpretations: the former is $4\pi \ell_B$ times the
local ionic charge density and the latter is $4\pi \ell_B q^2$ times
the local ionic concentration. The relations
Eqs.~(\ref{VarNetzP})--(\ref{VarNetzG}) are respectively similar in
form to the non-linear Poisson-Boltzmann (NLPB) and Debye-H\"uckel (DH)
equations, except that the charge and salt sources due to mobile
ions are replaced by their local values according to the Boltzmann
distribution.  On the one hand, Eq.~(\ref{VarNetzP}) is a
Poisson-Boltzmann like equation where appears the local charge
density proportional to $\sinh\phi_0$. This equation handles the
asymmetry induced by the surface through the electrostatic potential
$\phi_0$, which ensures electroneutrality. This asymmetry may be due
to the effect of the  surface charge on anion- and
cation-distributions (see Section~\ref{chargedfilm}) or due to
dielectric boundaries and image charges at neutral interfaces, which
give rise to interactions proportional to $q^2$, and induce a local
non-zero $\phi_0$ for asymmetric electrolytes (see
Section~\ref{waterair}). On the other hand, the generalized DH
equation Eq.~(\ref{VarNetzG}), where appears the local ionic
concentration proportional to $\cosh\phi_0$, fixes the Green's
function $v_0(\br,\br')$ evaluated at $\br$ with the charge source
located at $\br'$ and takes into account dielectric jumps at
boundaries.

These variational equations were first obtained within the
variational method by Netz and Orland~\cite{netz_var}. They were
also derived in Ref.~\cite{avdeev} within the Debye closure approach
and the BBGKY hierarchic chain. Yaroshchuk obtained an approximate
solution of the closure equations for confined electrolyte systems
in order to study ion exclusion from membranes~\cite{yarosch}.

Equations~(\ref{VarNetzP})--(\ref{VarNetzG}) enclose the limiting
cases of WC ($\Xi\to0$) and SC ($\Xi\to\infty$). To see that, it is
interesting to rewrite theses equations by renormalizing all lengths
and the fixed charge density, $\rho_s(\br)$, by the Gouy-Chapman
length according to $\tilde{\br}=\br/\ell_G$, $\tilde\rho_s(\tilde
\br)=\ell_G\rho_s(\br)/\sigma_s$ ($\sigma_s$ is the average surface
charge density). By introducing a new electrostatic potential
$\tilde\phi_0(\br)=q\phi_0(\br)$, one can express the same set of
equations in an adimensional form \bea
&&\tilde\Delta \tilde \phi_0(\tilde{\br})-\Lambda e^{-\frac{\Xi}{2}\tilde W(\tilde{\br})}\sinh\tilde{\phi}_0(\tilde{\br})=-2\tilde \rho_s(\tilde{\br})\label{VarNetzPXi}\\
&&-\tilde\Delta \tilde v_0(\tilde{\br},\tilde{\br}')+\Lambda
e^{-\frac{\Xi}{2}\tilde
W(\tilde{\br})}\cosh\tilde{\phi}_0(\tilde{\br})
\tilde v_0(\tilde{\br},\tilde{\br}')\nonumber\\
&&\hspace{5cm}=4\pi\delta(\tilde{\br}-\tilde{\br}')\label{VarNetzGXi}
\eea where $\tilde v_0=v_0\ell_G/\ell_B$, $\tilde W=W\ell_G/\ell_B$
and we have also introduced the rescaled fugacity
$\Lambda=8\pi\lambda\ell_G^3\Xi$~\cite{footnote1}. Now, one can
check that, in both limits $\Xi\to0$ and $\Xi\to\infty$, the
coupling between $\phi_0$ and $v_0$ in Eq.~(\ref{VarNetzP})
disappears and the theory becomes integrable. Finally, it is
important to note that this adimensional form of variational
equations allows one to focus on the role of $v_0(\br,\br)$ whose
strength is controlled by $\Xi$ in Eqs.~(\ref{VarNetzPXi})
and~(\ref{VarNetzGXi}). However, even at the numerical level, their
explicit coupling does not allow for exact solutions for general
$\Xi$.

In the present work, we make a restricted choice for $v_0(\br,\br')$
and replace the local salt concentration in the form of a local
Debye-H\"uckel parameter (or inverse screening length) $\kappa(\br)$
in Eq.~(\ref{VarNetzG}), \be \kappa(\br)^2=8\pi \ell_B q^2\lambda
e^{-\frac{q^2}{2}W(\br)}\cosh\left[q\phi_0(\br)\right], \ee by a
constant piecewise one $\kappa_v(\br)=\kappa_v$ in the presence of
ions and $\kappa_v(\br)=0$ in the salt-free parts of the system.
Note that it has been recently shown that many thermodynamic
properties of electrolytes are successfully described with a
Debye-H\"uckel kernel~\cite{janecek}.

The inverse kernel (or the Green's function) $v_0(\br,\br')$  is
then taken to be the solution to a generalized Debye-H\"uckel (DH)
equation \be\label{DH1}
\left[-\nabla(\epsilon(\br)\nabla)+\epsilon(\br)\kappa_v^2(\br)\right]v_0(\br,\br')=\frac{e^2}{k_BT}\delta(\br-\br')
\ee with the boundary conditions associated with the dielectric
discontinuities of the system \bea
\lim_{\br \to\Sigma^-}v_0(\br,\br')&=&\lim_{\br \to \Sigma^+}v_0(\br,\br'),\\
\lim_{\br \to \Sigma^-}\epsilon(\br)\nabla v_0(\br,\br')&=&\lim_{\br
\to \Sigma^+}\epsilon(\br)\nabla v_0(\br,\br') \label{limits_v} \eea
where $\Sigma$ denotes the dielectric interfaces. We now restrict
ourselves to planar geometries. We split the grand potential
(\ref{Fvar}) into three parts, $F_v=F_1+F_2+F_3$, where $F_1$ is the
mean electrostatic potential contribution, \bea\label{FvarIII}
F_1&=&S\int \mathrm{d}z\left\{-\frac{\left[\nabla\phi_0(z)\right]^2}{8\pi \ell_B}+\rho_s(z)\phi_0(z)\right.\nonumber\\
&&-\left.\sum_i\lambda_i
e^{-\frac{q_i^2}{2}W(z)-q_i\phi_0(z)}\right\}, \eea $F_2$ the kernel
part and $F_3$ the unscreened Van der Waals contribution. The
explicit forms of $F_2$ and $F_3$ are reported in Appendix
\ref{appendixVarFr}. The first variational equation is given by
$\partial F_v/\partial\kappa_v=\partial
\left(F_1+F_2\right)/\partial\kappa_v=0$. This equation is the
restricted case of Eq.~(\ref{VarNetzG}). As we will see below, its
explicit form depends on the confinement geometry of the electrolyte
system as well as on the form of $\epsilon(\br)$. The variational
equation for the electrostatic potential~\cite{footnote2} $\delta
F_v/\delta \phi_0(z)=0$ yields regardless of the confinement
geometry \be\label{eqvarI} \frac{\partial^2 \phi_0}{\partial
z^2}+4\pi \ell_B\rho_s(z) +\sum_i 4\pi\ell_B q_i\lambda_i
e^{-\frac{q_i^2}{2} W(z)-q_i\phi_0(z)}=0. \ee

The second-order differential equation~(\ref{eqvarI}), which is
simply the generalization of Eq.~(\ref{VarNetzP}) for a general
electrolyte in a planar geometry, does not have closed-form
solutions for spatially variable $W(z)$. In what follows, we
optimize the variational grand potential $F_v$ using restricted
forms for the electrostatic potential $\phi_0(z)$ and compare the
result to the numerical solution of Eq.~(\ref{eqvarI}) for single
interfaces and slit-like pores.

The single ion concentration is given by \be\label{rho}
\rho_i(z)=\lambda_i\;e^{-\frac{q_i^2}{2}W(z)-q_i\phi_0(z)} \ee and
its spatial integral by \be\label{rhoIn} \int
\mathrm{d}z\rho_i(z)=-\lambda_i\frac{\partial F_v}{\partial
\lambda_i}. \ee We define the Potential of Mean Force (PMF) of ions
of type $i$, $\Phi_i(z)$, as \be \Phi_i(z)\equiv
-\ln\frac{\rho_i(z)}{\rho_b} \label{defPMF} \ee By defining \be w(z)
\equiv W(z)-W_b \ee where $W_b$ is the value of $W(z)$ in the bulk
and comparing Eqs.~(\ref{rho})--(\ref{defPMF}), we find \bea
\Phi_i(z) &=& \frac{q_i^2}2w(z)+q_i\phi_0(z)\label{PMF2}\\
\frac{q_i^2}2W_b &=& \ln\gamma^b_i\equiv\mu_i
-\ln(\rho_b\lambda_t^3)\label{Wb} \eea
Hence, $q_i^2W_b/2$  is nothing else but the excess chemical potential of ion $i$ in the bulk and $q_i^2W(z)/2=\ln\gamma_i(z)$ is its generalization  for ion $i$  at distance $z$ from the interface. They are related to the activity coefficients $\gamma_i^b$ and $\gamma_i(z)$. Note that the zero of the chemical potential is fixed by the condition that $\phi_0$ vanishes in the bulk. The PMF, Eq.~(\ref{PMF2}), is thus the mean free energy per ion (or chemical potential) needed to bring an ion from the bulk at infinity  to the point at distance $z$ from the interface, taking into account correlations with the surrounding ionic cloud.\\

Before applying the variational procedure to single and double
interfaces, let us consider the variational approach in the bulk. In
this case, the variational potential $\phi_0$ is equal to 0, and the
variational grand potential $F_v$ only depends on $\kappa_v$. Two
minima appear: one metastable minimum $\kappa_v^0$ at low values of
$\kappa_v$, and a global minimum at infinity  ($F_v\to-\infty$ for
$\kappa_v\to\infty$) which is unphysical since at these large
concentration values, finite size effects should be taken into
account. It has been shown by introducing a cutoff at small
distances~\cite{curtis}, that, for physical temperatures, this
instability disappears and the global minimum of $F_v$ is
$\kappa_v^0\approx\kappa_b$ given by the Debye--H\" uckel limiting
law: \be \left\lbrace
\begin{array}{l}
\mu_i=\ln(\rho_b\lambda_t^3)-\frac{q_i^2}2\kappa_b\ell_B\\
\kappa_b^2=4\pi \ell_B \sum_i q_i^2\rho_{i,b}
\end{array}\right.
\label{DH} \ee From Eq.~(\ref{Wb}), we thus find
$W_b\approx-\kappa_b\ell_B$ and the potential $w(z)$ reduces to
\be\label{PMF} w(z)\approx v_0(z,z)-v_c^b(0)+\kappa_b\ell_B, \ee
which will be adopted in the rest of the paper. Furthermore,
problems due to the formation of ion pairs do not enter at the level
of the variational approach we have adopted. Let us also report the
following conversion relations \bea\label{CONV}
I_b&\simeq&  0.19(\kappa_b\ell_B)^2\hspace{1mm}\mbox{mol.L}^{-1}\\
\kappa_b&\simeq& 3.29\sqrt{I}\hspace{1mm}\mbox{nm}^{-1},
\nonumber \eea
where $I=1/2\sum_iq_i^2\rho_{i,b}$ is the ionic
strength expressed in $\mbox{mol.L}^{-1}$. Finally, the single-ion
densities are given by \be
\rho_i(z)=\rho_{i,b}e^{-\frac{q_i^2}{2}w(z)-q_i\phi_0(z)}. \ee

\section{Single interface}
\label{singleinter}

\begin{figure}
\includegraphics[width=.9\linewidth]{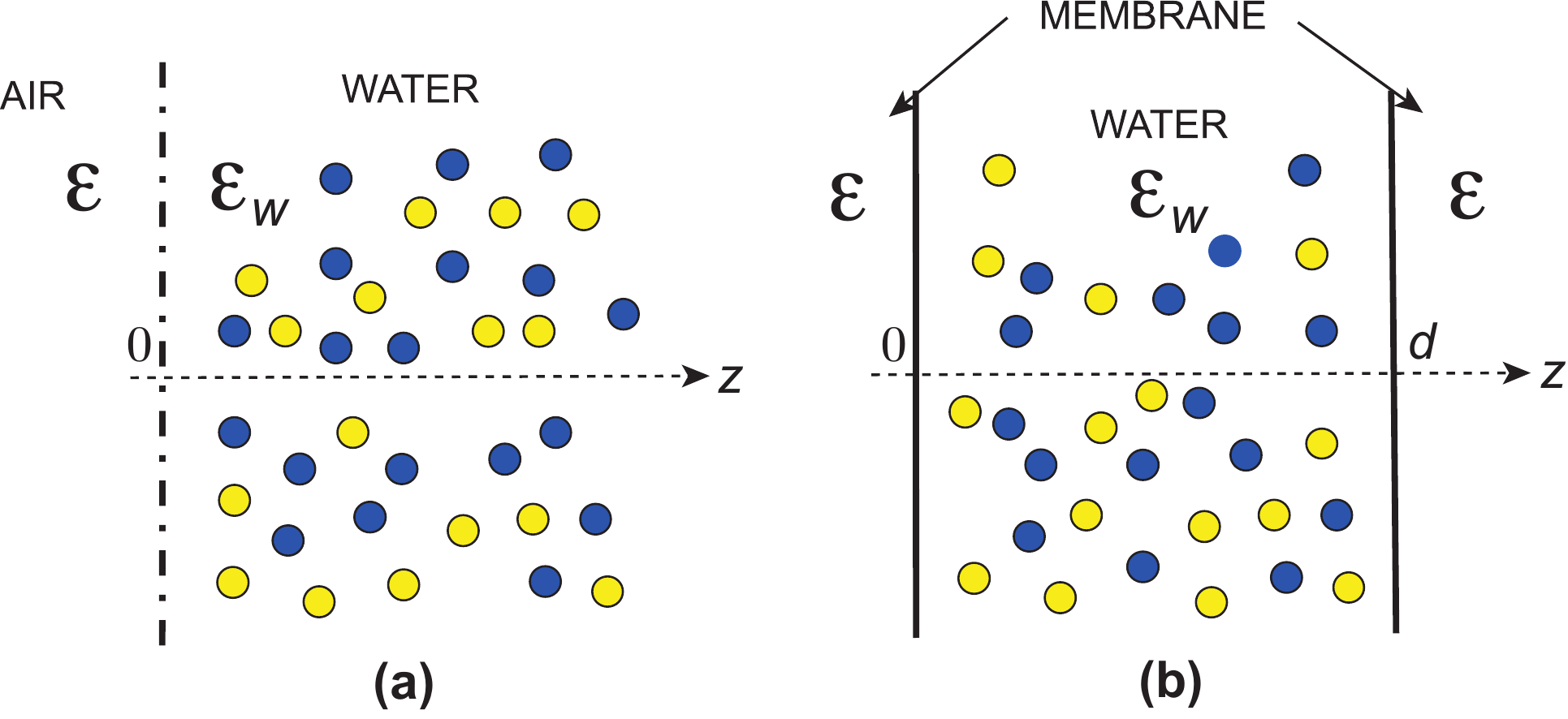}
\caption{(color online) Geometry for a single dielectric interface
(e.g. water--air) (a) and double interfaces or slit-like pores (b).}
\label{sketch}
\end{figure}
The single interfacial system considered in this Section consists of
a planar interface separating a salt-free left half-space from a
right half-space filled up with an electrolyte solution of different
species (Fig.~\ref{sketch}). In the general case, the dielectric
permittivity of the two half spaces may be different (we note
$\epsilon$ the permittivity in the salt-free part). The Green's
function, which is chosen to be solution of the DH equation with
$\epsilon(z)=\epsilon\theta(-z)+\epsilon_w\theta(z)$ and
$\kappa(z)=\kappa_v\theta(z)$ where $\theta(z)$ stands for the
Heaviside distribution, is given for $z>0$ by~\cite{bravina}
\bea\label{CHPO}
w(z)&=& \ell_B(\kappa_b-\kappa_v)\\
&&+\ell_B\int_0^\infty
\frac{k\mathrm{d}k}{\sqrt{k^2+\kappa_v^2}}\Delta(k/\kappa_v)\;e^{-2\sqrt{k^2+\kappa_v^2}
z}\nonumber \eea where \be
\Delta(x)=\frac{\epsilon_w\sqrt{x^2+1}-\epsilon
x}{\epsilon_w\sqrt{x^2+1}+\epsilon x}. \label{defDelta} \ee and
$F_2$ [Eq.~(\ref{FvarIV})] can be analytically
computed~\cite{curtis} \be\label{FGI}
F_2=V\frac{\kappa_v^3}{24\pi}+S\Delta\frac{\kappa_v^2}{32\pi}, \ee
where \be
\Delta\equiv\Delta(x\to\infty)=\frac{\epsilon_w-\epsilon}{\epsilon_w+\epsilon}.
\label{jump} \ee The first term on the rhs. of Eq.~(\ref{FGI}) is
simply the volumic Debye free energy associated with a hypothetic
bulk with a Debye inverse screening length $\kappa_v$ and the second
term on the rhs. involves interfacial effects, including the
dielectric jump $\Delta$, and $\kappa_v$.

For the single interface system, as seen in Section~\ref{varcalc},
$F_3$ is independent of $\kappa_v$ and $\phi_0(z)$, which means that
it does not contribute to the variational equations. By minimizing
Eqs.~(\ref{FvarIII}) and~(\ref{FGI}) with respect to $\kappa_v$ for
fixed $\phi_0(\br)$ and taking $V\to \infty$, one exactly find the
same variational equation for $\kappa_v$ as for the bulk case.
Hence, as discussed above, we have $\kappa_v=\kappa_b$ given by
Eq.~(\ref{DH}) and the first term of the rhs. of Eq. (\ref{CHPO})
vanishes.
\begin{figure}[t]
\includegraphics[width=.8\linewidth]{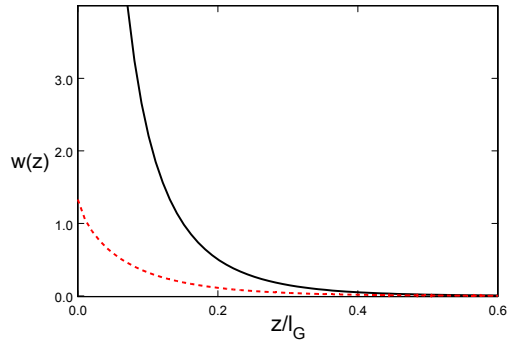}
\caption{(color online) Potential $w(z)$ in units of $k_BT$ for
$\epsilon=0$ (black solid curve), Eq.~(\ref{PotII}), and
$\epsilon=\epsilon_w$ (red dashed curve) and $\kappa_b\ell_B=4$.}
\label{Green}
\end{figure}
This result was obtained in~\cite{curtis} for the special case
$\phi(\br)=0$. It is of course not surprising to end up with the
same result for finite $\phi(\br)$ since we know that the
electrostatic potential should vanish in the bulk. This potential
combines in an intricate way both the image charge and solvation
contributions due to the presence of the interface. The image force
corresponds to the interaction of a given ion with the polarized
charges at the interface and is equivalent to the interaction of the
charged ion with its image located at the other side of the
dielectric surface. As it is well known, the image charge
interaction is repulsive for $\epsilon<\epsilon_w$ (e.g. water-air
interface) and attractive for $\epsilon>\epsilon_w$ (the case for an
electrolyte-metal interface)~\cite{jackson_book}. The interfacial
reduction in solvation arises because an ion always prefers to be
screened by other ions in order to reduce its free energy. Hence, it
is attracted towards areas where the ion density is maximum (at
least at not too high concentrations for which steric repulsion may
predominate). This term is non-zero even for $\epsilon=\epsilon_w$
since for an ion close to the interface, there is a ``hole'' of
screening ions in the salt-free region (where $\kappa_v=0$).
Although our choice of homogeneous variational inverse screening
length allows us to handle the deformation of ionic atmospheres near
interfaces that are impermeable to ions, it does not allow us to
treat in detail the local variations in ion solvation free energy
arising from ion-ion correlations (except in an average way in
confined geometries where $\kappa_v$ can differ from the bulk
value of the inverse screening length, see Section IV below).

Equation (\ref{CHPO}) simplifies in three cases :

\noindent1) For $\epsilon=0$ ($\Delta=1$), where the solvation
effect vanishes because the lines of forces are totally excluded
from the air region~\cite{jackson_book}, Eq.~(\ref{CHPO}) reduces to
\be\label{PotII} w_0(z)=\ell_B\frac{e^{-2\kappa_bz}}{2z}. \ee This
is the case where the image charge repulsion is the strongest (see
Fig.~\ref{Green}).

\noindent2)A slightly better approximation for $\epsilon\neq0$ can
be obtained by artificially allowing salt to be present in the air
region. This gives rise to the ``undistorted ionic atmosphere''
approximation~\cite{yarosch}, for which $w(z)$ in Eq.~(\ref{PotII})
is multiplied by $\Delta$: \be\label{PotIII} w(z)=
\Delta\ell_B\frac{e^{-2\kappa_bz}}{2z}. \ee Solvation effects are
now absent and salt exclusion arises solely from dielectric
repulsion. Eq.~(\ref{PotIII}) is exact for arbitrary $\kappa_b$ and
$\Delta=1$, or arbitrary $\Delta$ and $\kappa_b=0$.

\noindent3)In the absence of a dielectric discontinuity
$\epsilon=\epsilon_w$ ($\Delta=0$), the potential can be expressed
as \bea\label{PotIV}
w(z)&=&\kappa_b\ell_Bf(\kappa_bz)\\
f(x)&=&\frac{(1+x)^2e^{-2x}}{2x^3}-\frac{K_2(2x)}{x}\nonumber \eea
where $K_2(x)$ is the Bessel function of the second kind. One
notices that unlike the case $\Delta>0$, the potential has a finite
value at the interface, i.e. $w(0)=\kappa_b\ell_B/3$.

We note that in this case of one interface, we have
$\lim_{z\to\infty}\phi_0(z)=0 $ and the fugacity $\lambda_i$ of each
species is fixed by its bulk concentration according to
\be\label{CHP} \rho_{i,b}=\lim_{z\to\infty}\rho_i(z)=\lambda_i
e^{\frac{q_i^2}{2}\kappa_b \ell_B} \ee where we used
Eq.~(\ref{rho}).

\subsection{Neutral dielectric interface}
\label{waterair}

\begin{figure}[t]
(a)\includegraphics[width=.95\linewidth]{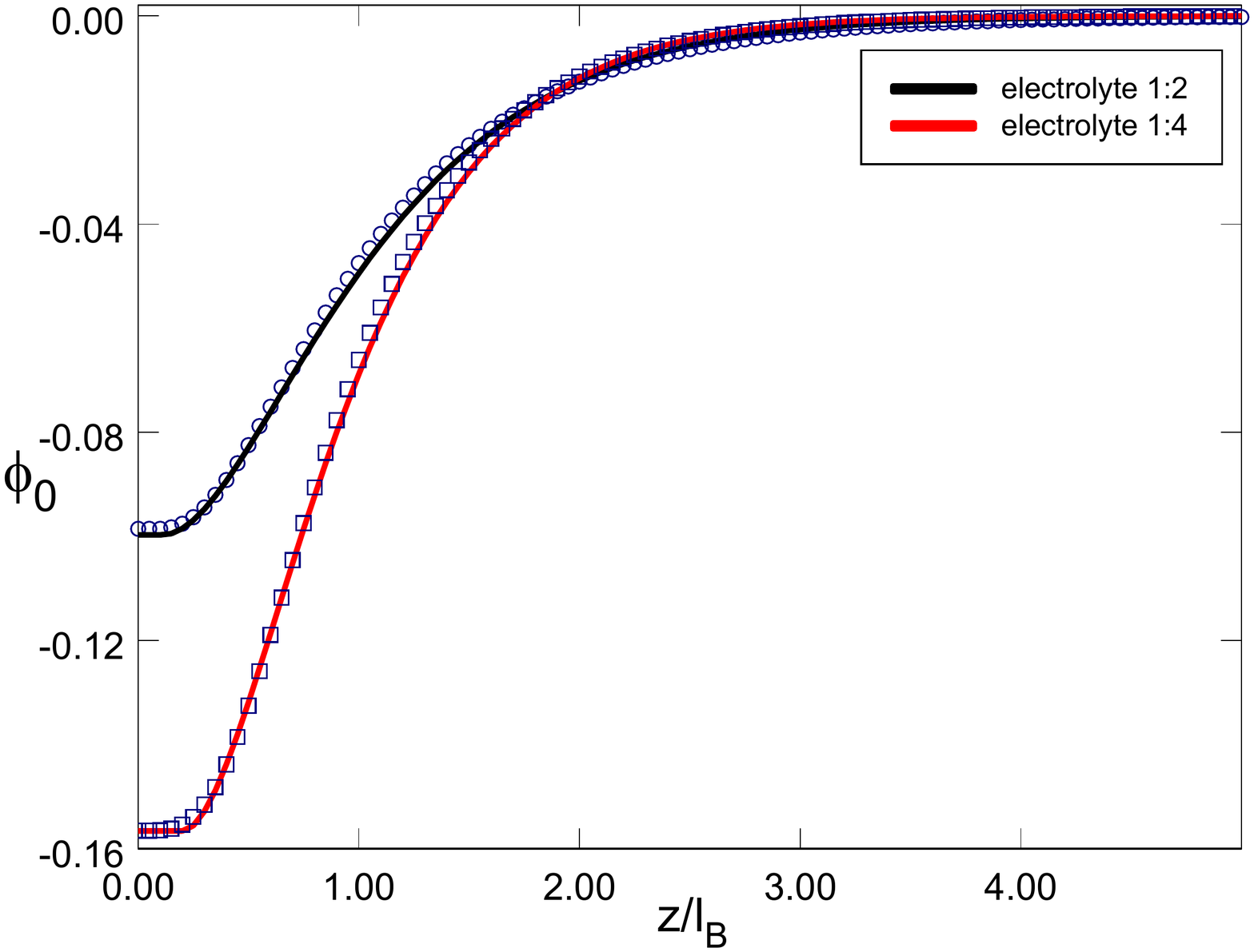}
(b)\includegraphics[width=.9\linewidth]{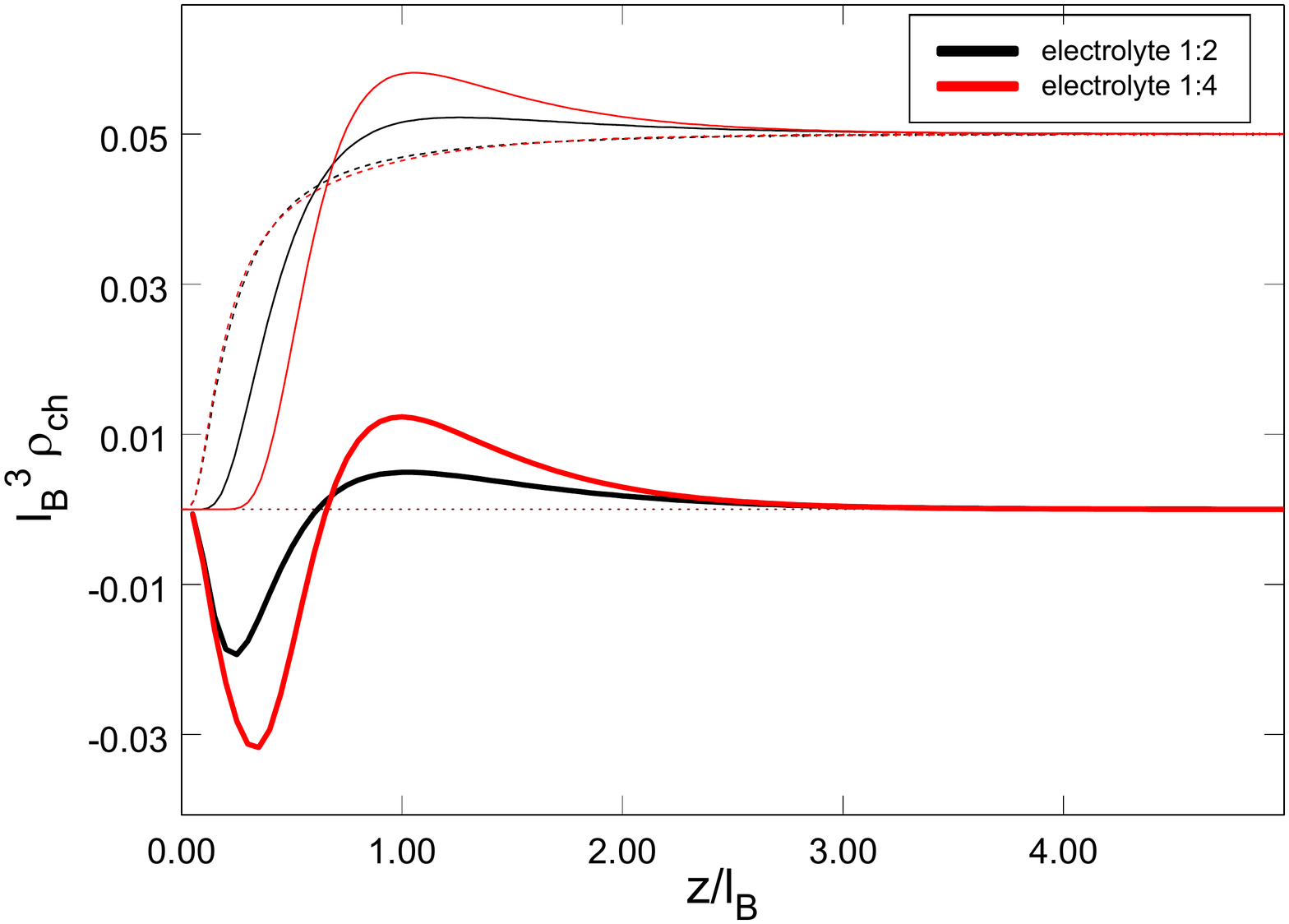}
\caption{(color online) (a) Electrostatic potential $\phi_0$ (in
$k_BT$ units) for asymmetric electrolytes: numerical solution of
Eq.~(\ref{eqvarII}) (symbols) and variational choice,
Eq.~(\ref{TriPotWA}) (solid lines), for divalent and quadrivalent
ions and $\rho_-=0.242$ M. Variational parameters are
$\kappa_\phi\simeq1.4\kappa_b$, $a/\ell_B=0.12;0.21$ and
$\varphi=-0.10;-0.156$. (b)~Associated local charge density profile
(thick lines) and anion (dashed lines) and cation (thin solid
lines) concentrations.} \label{FieldAsym}
\end{figure}

We investigate in this section the physics of an asymmetric
electrolyte close to a neutral dielectric interface (e.g.
water--air, liquid--liquid or liquid--solid interface) located at
$z=0$ ($\sigma_s=0$). For the sake of simplicity, we assume
$\epsilon=0$, which is a very good approximation for the air-water
interface characterized by $\epsilon=1$ (see the discussion in
Ref.~\cite{onsager}). Hence we keep the approximation  $w(z)=w_0(z)$
given by Eq.~(\ref{PotII}). The electrolyte is composed of two
species of bulk density $\rho_+$ and $\rho_-$ and charge
$(q_+e)$,$-(q_-e)$ with $q_+>q_-$. In order to satisfy the
electroneutrality in the bulk, we impose $\rho_+q_+=\rho_-q_-$.
According to Eq.~(\ref{DH}), the bulk inverse screening length noted
$\kappa_b$ is given by \be \kappa^2_b=4\pi
\ell_Bq_-\rho_-\left(q_-+q_+\right)\label{SCasym} \ee and the
variational equation~(\ref{eqvarI}) for the electrostatic potential
is a modified Poisson-Boltzmann equation \be \frac{\partial^2
\phi_0}{\partial z^2}+4\pi \ell_B\rho_{\rm ch}(z)=0\label{eqvarII}
\ee with a local charge concentration \be \rho_{\rm
ch}(z)=\rho_-q_-\left[e^{-\frac{q_+^2}{2}w(z)-q_+\phi_0(z)}-e^{-\frac{q_-^2}{2}w(z)+q_-\phi_0(z)}\right].
\ee Equation~(\ref{eqvarII}) can not be solved analytically. Its
numerical solution, obtained using a 4th order Runge-Kutta method,
is plotted in Fig.~\ref{FieldAsym}(a) for asymmetric electrolytes
with divalent and quadrivalent cations and the local charge density
is plotted in Fig.~\ref{FieldAsym}(b).

Fig.~\ref{FieldAsym} clearly shows that, very close to the
dielectric interface for $z<a$, image charge repulsion expulses all
ions (since $\rho_{\rm ch}(z)\sim\exp(-1/z)$ has an essential
singularity) and $\phi_0$ is flat. For $z>a$,  but still close to
the interface, there is a layer where the electrostatic field is
almost constant ($\phi_0$ increases linearly), which is created by
the charge separation of ions of different valency due to repulsive
image interactions. The intensity of image forces increases with the
square of ion valency and close to the interface, $\rho_{\rm
ch}(z)<0$ since we assumed $q_+>q_-$ (the case for $\mbox{MgI}_2)$.
To ensure electroneutrality, the local charge then becomes positive
when we move away from the surface (Fig.~\ref{FieldAsym}(b)), and
the electrostatic potential goes exponentially to zero with a
typical relaxation constant $\kappa_\phi$. Moreover, in
Fig.~\ref{FieldAsym}(a) one observes that when the charge asymmetry
increases, the electrostatic potential also increases. Knowing that
for symmetric electrolytes, $\phi_0=0$, our results confirm that the
charge asymmetry is the source of the electrostatic potential
$\phi_0$. Fig.~\ref{FieldAsym}(b) is qualitatively similar to Fig.~1
of Bravina who had derived an integral solution of
Eq.~(\ref{eqvarII}) by using an approximation valid for
$\kappa_b\ell_B\ll 1$~\cite{bravina}.
In order to go further in the description of the interfacial
distribution of ions, we look for a restricted variational function
$\phi_0(z)$ which not only contains a small number of variational
parameters (such as $a$ and $\kappa_\phi$) but also is as close as
possible to the numerical solution. As suggested by the description
of Fig.~\ref{FieldAsym}, a continuous piecewise $\phi_0(z)$  is
necessary to account for the essential singularity of $\rho_{\rm
ch}(z)$. To show this, let us expand Eq.~(\ref{eqvarII}) to order
$\phi_0$: \bea \frac{\partial^2 \phi_0}{\partial z^2} &\approx&
-4\pi \ell_B\rho_-q_-\left[e^{-\frac{q_+^2}{2}w(z)}-
e^{-\frac{q_-^2}{2}w(z)}\right]\label{eqvarTY}\\
&& +4\pi
\ell_B\rho_-q_-\left[q_+e^{-\frac{q_+^2}{2}w(z)}+q_-e^{-\frac{q_-^2}{2}w(z)}\right]\phi_0.\nonumber
\eea This linearization is legitimate, as seen in
Fig.~\ref{FieldAsym}: $q_+|\phi_0(z)|<1$ is satisfied for physical
valencies. The first term in the rhs. of Eq.~(\ref{eqvarTY})
corresponds to an effective local charge source while the second
term is responsible for the screening of the potential. If we
observe the charge distribution for $q^2w(z)>q\phi_0$(z) and $z>a$,
i.e. the first term of the rhs. of Eq.~(\ref{eqvarTY}), we notice
that it  behaves like a distorted peak. The simplest function having
a similar behavior is $f(z)=cze^{-\kappa_\phi z}$, where $c$ and
$\kappa_\phi$ are constants. Hence, we choose a restricted
variational piecewise solution $\phi_0(z)$ \be\label{TriPotWA}
\phi_0(z)=\left\lbrace
\begin{array}{ll}
\varphi & \mathrm{for}\quad z\leq a,\\
\varphi\left[1+\kappa_\phi (z-a)\right]e^{-\kappa_\phi(z-a)} &
\mathrm{for}\quad z\geq a.
\end{array}\right.
\ee whose derivation is explained in Appendix \ref{appendixWA}. The
variational parameters are the constant potential $\varphi$, the
depletion distance $a$ and the inverse screening length
$\kappa_\phi$. The grand potential (\ref{FreeWA}) derived for this
solution was optimized with respect to the variational parameters
using the Mathematica software. The restricted variational
potential~(\ref{TriPotWA}) is compared to the numerical solution of
Eq.~(\ref{eqvarII}) in Fig.~\ref{FieldAsym} for electrolytes $1:2$
and $1:4$ and $\rho_-\ell_B^3=0.05$. The agreement is excellent. One
notices that the screening of the effective surface charge created
by dielectric exclusion enters into play when $z>\kappa_\phi^{-1}$.
Finally, let us note that since $\kappa_bl_B=1.37$ and
$\kappa_bl_B=1.77$ respectively for the monovalent and quadrivalent
electrolytes in Fig.~\ref{FieldAsym}, the method adopted by Bravina
is not valid.

To summarize, the charge separation is taken into account by the
potential $\varphi$ (which increases with $q_+/q_-$) and the
relaxation constant $\kappa_\phi\simeq1.4\kappa_b$ is almost
independent of $q_+/q_-$. Interestingly, the variational parameter
$a/\ell_B\simeq0.1-0.2$ is less than 1~nm. Indeed, for finite size
ions, $w(z)$ differs from Eq.~(\ref{PotII}) very close to the
interface and reaches a finite value at $z=0$. The size of this
region exactly corresponds to $a$ which is of the order of an ion
radius. This is thus an artifact of our point-like ion model and
occurs only for asymmetric electrolytes at neutral surfaces.

\begin{figure}[t]
\includegraphics[width=.9\linewidth]{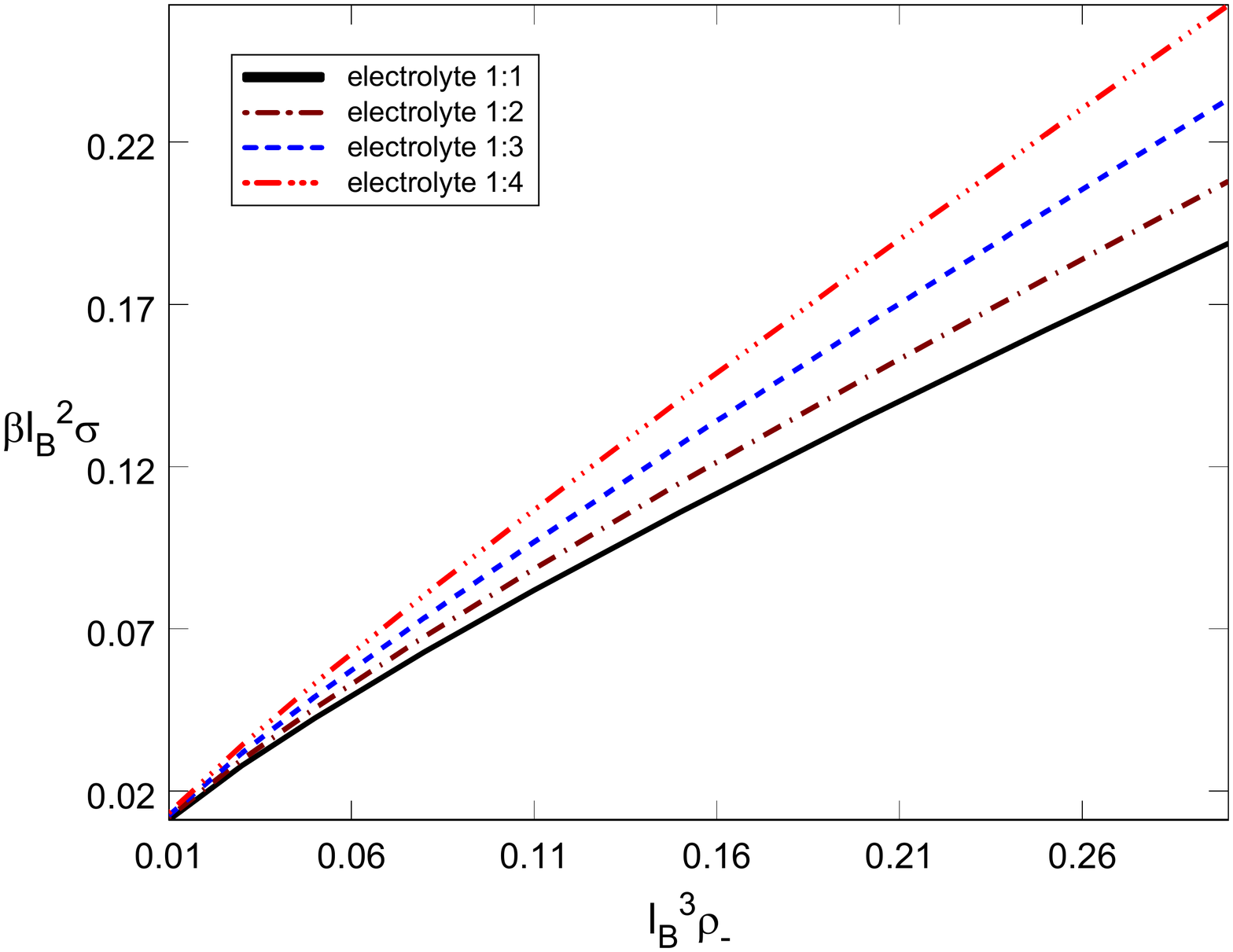}
\caption{(color online) Surface tension $\ell_B^2\sigma/k_BT$ for
asymmetric electrolytes vs. the anion bulk concentration, for
increasing asymmetry $q_+/q_-=1$ to 4 from bottom to top.}
\label{SurfaceTen}
\end{figure}

The surface tension $\sigma$ is equal to the excess grand potential
defined as the difference between the grand potential of the
interfacial system and that of the bulk system: \bea
\sigma&=&\frac{\Delta\kappa_b^2}{32\pi}-\frac{\kappa_\phi\varphi^2}{32\pi \ell_B}\label{STWA}\\
&& - \rho_-\int_0^\infty  \mathrm{d}z \left\{\left[e^{-\frac{q_-^2}{2}w(z)+q_-\phi_0(z)}-1\right] \right.\nonumber\\
&&
+\left.\frac{q_-}{q_+}\left[e^{-\frac{q_+^2}{2}w(z)-q_+\phi_0(z)}-1\right]\right\}.\nonumber
\eea The surface tension for electrolytes characterized by $q_-=1$
and $q_+=1$ to $4$ is plotted in Fig.~\ref{SurfaceTen} as a a
function of $\rho_-$, because the anion density is an experimentally
accessible parameter. Unlike symmetric electrolytes~\cite{curtis}, a
plot with respect to $\kappa_b^2$ may lead to a different behavior.
One notices that the increase in valency asymmetry leads to an
important increase of the surface tension. This is of course mainly
due to the reduction of the cation density in the bulk by a factor
of $q_-/q_+$ necessary to satisfy the bulk electroneutrality (see
the second term in the integral of Eq.~(\ref{STWA})).

\subsection{Charged surfaces}
\label{chargedfilm}

We now consider a symmetric electrolyte in the proximity of an
interface of constant surface charge $\sigma_s<0$ located at $z=0$.
The variational equation~(\ref{eqvarI}) simplifies to
\be\label{eqvarV} \frac{\partial^2 \tilde\phi_0}{\partial
\tilde{z}^2}=2\delta(\tilde{z})+\tilde{\kappa}_b^2
e^{-\frac{\Xi}{2}\tilde w(\tilde{z})}\sinh\tilde\phi_0. \ee The
mean-field limit ($\Xi\to0$) of this equation corresponds to the
NLPB equation, whose solution reads \be\label{NLPBMF}
\tilde\phi_0(\tilde z)=4\mathrm{arctanh}\left(\gamma_b
e^{-\tilde{\kappa}_b\tilde{z}}\right) \ee where
$\gamma_b=\tilde{\kappa}_b-\sqrt{1+\tilde{\kappa}_b^2}$. In this
Section, we show that a piecewise solution for the electrostatic
potential similar to the one introduced in Section~\ref{waterair}
agrees very well with the numerical solution of Eq. (\ref{eqvarV}).
Inspired by the existence of a salt-free layer close to the
interface and a mean-field regime far from the interface (WC), we
propose two types of piecewise variational functions (see Appendix
\ref{appendixIC}). The first variational choice obeys the Poisson
equation in the first zone of size $h$ and the non-linear
Poisson-Boltzmann solution in the second zone : \be\label{resNL}
 \tilde\phi_0^{\rm NL}(\tilde{z})=\left\lbrace
 \begin{array}{ll}
 4\mathrm{arctanh}\gamma+2(\tilde{z}-\tilde{h})&\mathrm{for}\quad\tilde{z}\leq \tilde{h},\\
 4\mathrm{arctanh}\left(\gamma e^{-\tilde{\kappa}_\phi(\tilde{z}-\tilde{h})}\right)&\mathrm{for}\quad \tilde{z}\geq \tilde{h},
 \end{array}\right.
\ee where
$\gamma=\tilde{\kappa}_\phi-\sqrt{1+\tilde{\kappa}_\phi^2}$.
Variational parameters are $h$ and an effective inverse screening
length $\kappa_\phi$. The second type of trial potential obeys the
Laplace equation with a charge renormalization in the first zone and
the linearized Poisson-Boltzmann solution in the second zone : \be
 \tilde\phi_0^{\rm L}(\tilde{z})=\left\lbrace
 \begin{array}{ll}
 -\frac{2\eta}{\tilde{\kappa}_\phi}+2\eta(\tilde{z}-\tilde{h})& \mathrm{for}\quad\tilde{z}\leq\tilde{h},\\
 -\frac{2\eta}{\tilde{\kappa}_\phi}e^{-\tilde{\kappa}_\phi(\tilde{z}-\tilde{h})}&\mathrm{for}\quad \tilde{z}\geq\tilde{h}.
 \end{array}\right.
 \label{resL}
\ee Variational parameters  are $\tilde h$, $\tilde\kappa_\phi$, and
the charge renormalization $\eta$, which takes into account the
non-linear effects at the mean-field level~\cite{netz_var}.
\begin{figure}
\includegraphics[width=.9\linewidth]{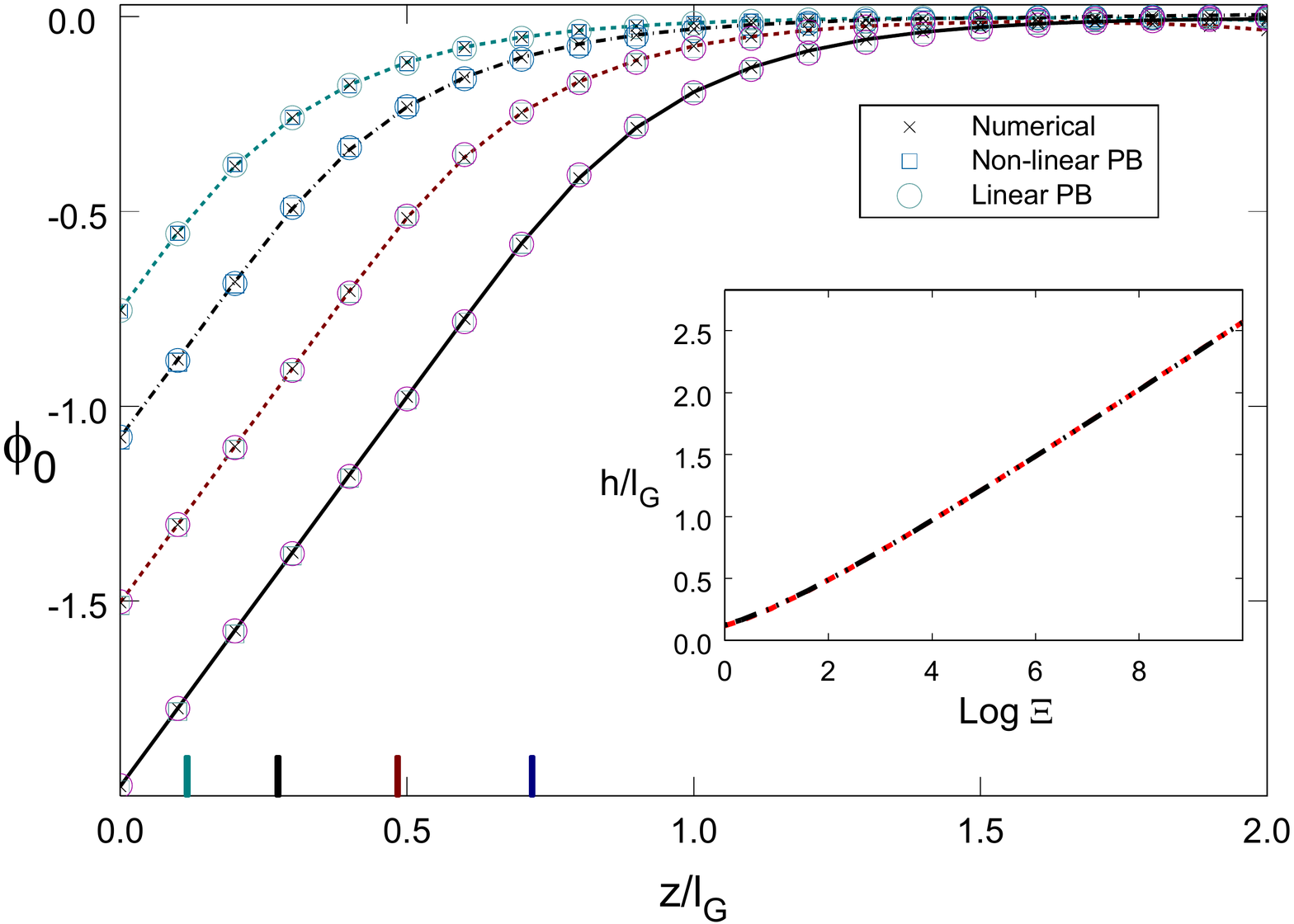}
\caption{(color online) Electrostatic potential, $\phi_0$ (in units
of $k_BT$): numerical solution of Eq.~(\ref{eqvarV}) (symbols) and
restricted variational choices Eqs.~(\ref{resNL}) and~(\ref{resL})
for $\epsilon=0$, $\kappa_{b}\ell_G=4$, and $\Xi=1,10,100$, and 1000
(from top to bottom). The variational parameters are respectively
$\kappa_\phi=3.83,3.74,3.69,3.66$ and $\eta\simeq1$. Markers on the
$x$-axis denote, for each curve, the size, $\tilde h$, of the SC
zone, plotted vs. $\ln\Xi$ in the Inset.} \label{FieldSym}
\end{figure}
The explicit form of the associated variational free energies are
reported in Appendix \ref{appendixIC}. The inset of
Fig.~\ref{FieldSym} displays the size of the SC layer $h$ against
$\Xi$. Our approach predicts a logarithmic dependence
$\tilde{h}\propto \ln\Xi$, the factor behind the logarithm being
$\tilde{\kappa}_b^{-1}$ for $\tilde{\kappa}_b\gg1$. The restricted
choices for $\phi_0$ are compared with the full numerical solution
of Eq.~(\ref{eqvarII}) in the same figure for $\epsilon=0$. We see
that, as in the previous section, the numerical solution and the
restricted ones match perfectly. Hence salt-exclusion effects are
essentially carried by the parameter $h$. Furthermore, one notices
that $\tilde\phi_0(\tilde{z})$ relaxes to zero between
$\tilde{z}=\tilde{h}$ and
$\tilde{z}=\tilde{h}+2\tilde{\kappa}_\phi^{-1}$. At
$\kappa_b\ell_G=4$ we are in the linear regime of the PB equation
and therefore  one has $\eta\simeq1$. The charge renormalization
idea was introduced by Alexander \textit{et al.}~\cite{Alex}, who
showed that the non-linearity of the PB equation can be effectively
taken into account at long distances by renormalizing the fixed
charge source and extending the linearized zone where
$|\tilde\phi_0|<1$ to the whole domain. A linear solution of the
form Eq.~(\ref{resL}) can be very helpful for complicated geometries
or in the presence of a non-uniform charge distribution where the
NLPB equation does not present an analytical solution even
at the mean-field level. These issues will be discussed in a future
work.
\begin{figure}
(a)\includegraphics[width=.8\linewidth]{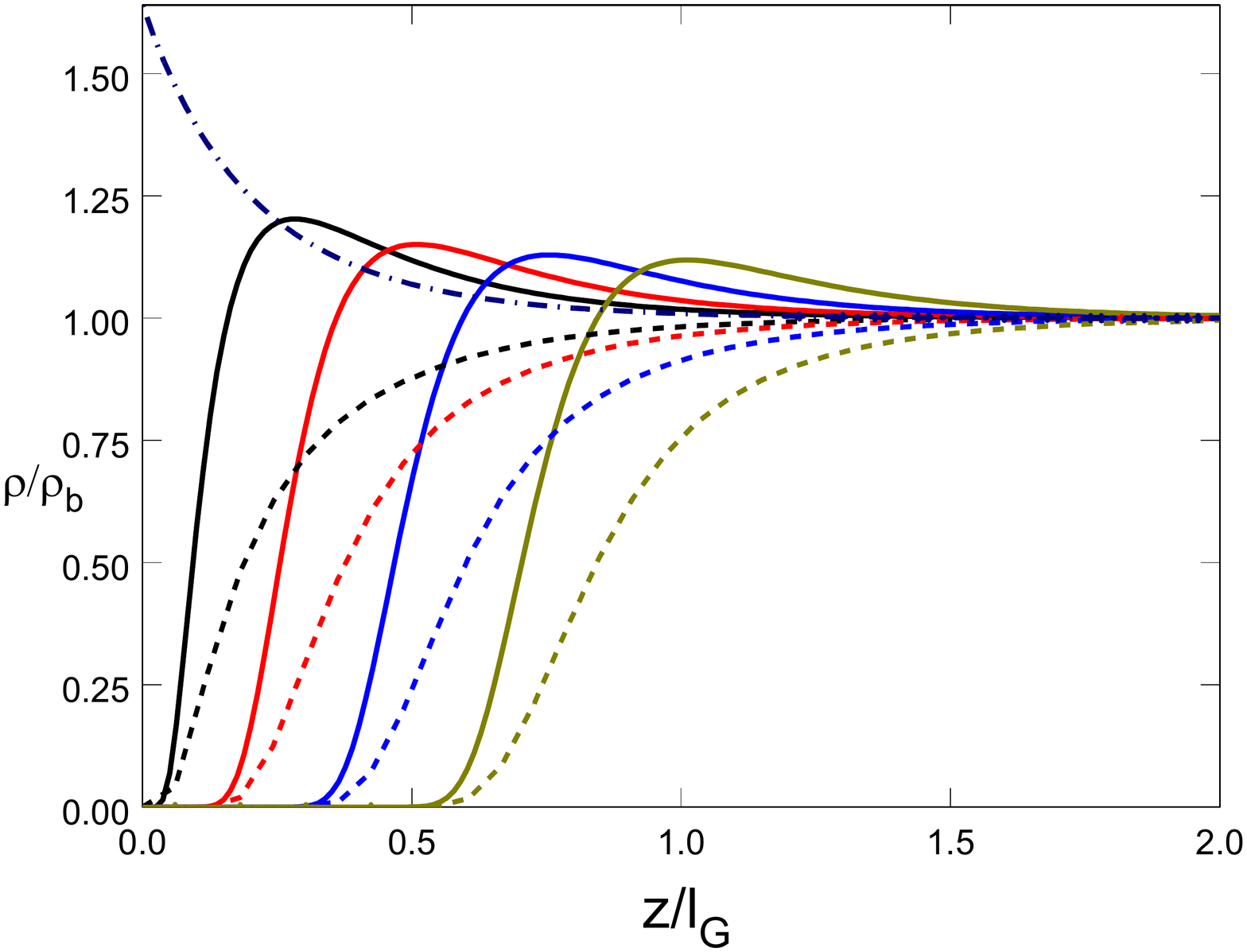}
(b)\includegraphics[width=.8\linewidth]{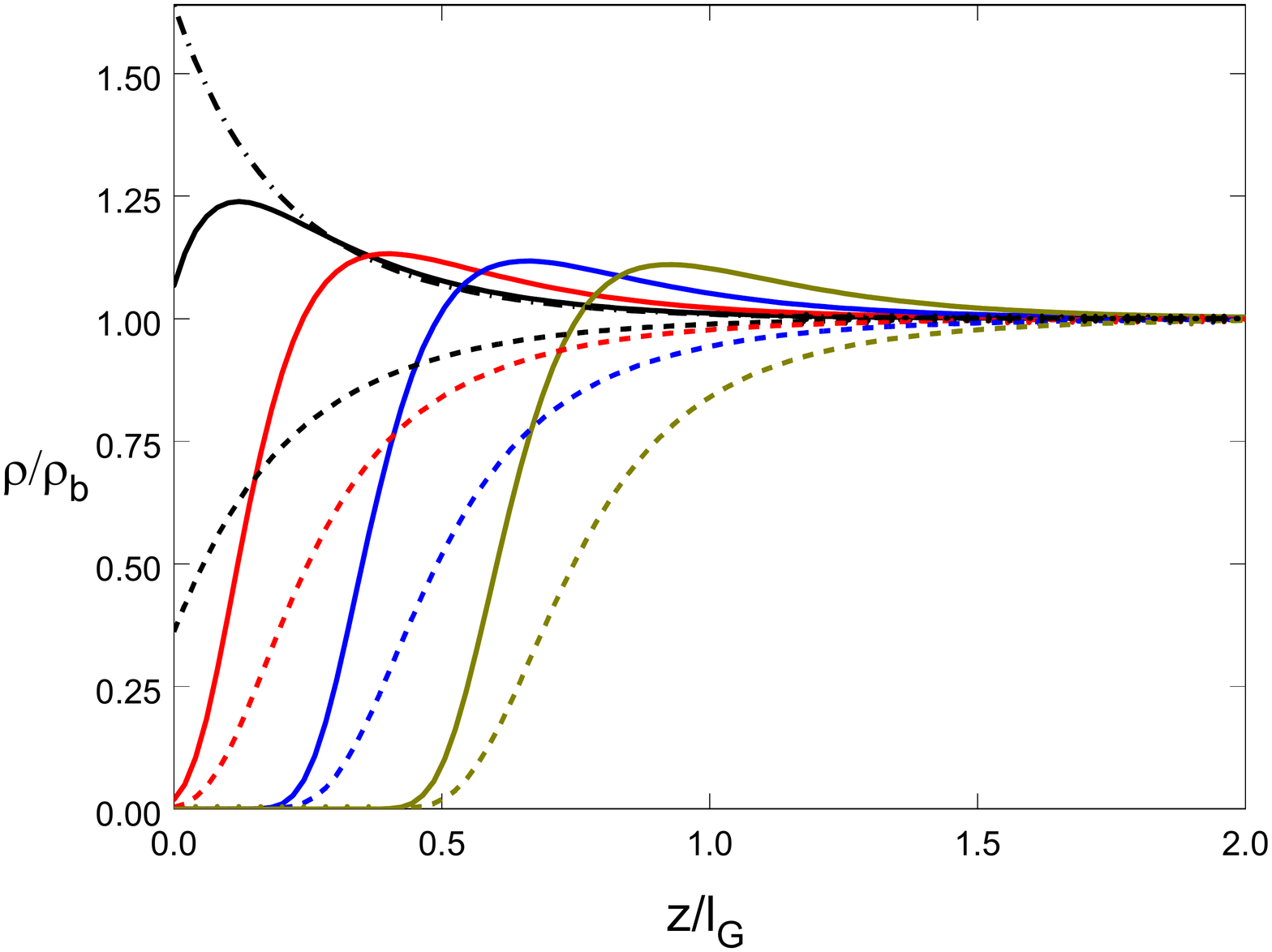} \caption{(color
online) Ion densities for $\kappa_{b}\ell_G=4$, and (a) $\epsilon=0$
and (b) $\epsilon=\epsilon_w$, for increasing coupling parameter:
from left to right, $\Xi=1,10,100$, and 1000. Solid lines correspond
to counterions, dashed lines to coions and dashed-dotted lines to
the Poisson-Boltzmann result~(\ref{NLPBMF}).} \label{IONdenSym}
\end{figure}

Figure~\ref{IONdenSym} displays the ion concentrations $\rho_i
(z)/\rho_{i, b} = e^{-\Phi_i}$, which are related to the ion PMF
Eq.~(\ref{defPMF}), computed with the restricted solution
Eq.~(\ref{resNL}) for several values of $\Xi$. As already said in
the Introduction, in rescaled distance, the coupling parameter $\Xi$
measures the strength of the excess chemical potential, $w(z)$. We
first see that for coions as well as for counterions, the depletion
layer in rescaled units in the proximity of the dielectric interface
increases with $\Xi$ due to the image charge repulsion and/or
solvation effect, i.e. the term $e^{-\frac{\Xi}{2}\tilde
w(\tilde{z})}$ in Eq.~(\ref{eqvarV}). Furthermore, one notices that
the counterion density exhibits a maximum. This concentration peak
is due to the competition between the attractive force towards the
charged wall and the repulsive image and solvation interactions. It
is important to note that in the particular case
$\epsilon=\epsilon_w$, there is no depletion layer for $\Xi<10$.

\section{Double Interface}
\label{dbleinter}

In this Section, the variational method is applied to a double
interface system which consists of a slit-like pore of thickness
$d$, in contact with an external ion reservoir at its extremities
(Fig.~\ref{sketch}). The dielectric constant is $\epsilon_w$ inside
the pore and $\epsilon$ in the outer space. The electrolyte occupies
the pore and the external space is salt-free. The solution of the DH
equation~(\ref{DHFour}) in this geometry is~\cite{yarosch} \bea
w(z)&=&(\kappa_b-\kappa_v)\ell_B\label{GreenConf}\\
&+&\ell_B\int_0^\infty\frac{k\mathrm{d}k}{\sqrt{k^2+\kappa_v^2}}\frac{\Delta(k/\kappa_v)}{e^{2d\sqrt{k^2+\kappa_v^2}}-\Delta^2(k/\kappa_v)}\nonumber\\
&\times&\left[2\Delta(k/\kappa_v)+e^{2(d-z)\sqrt{k^2+\kappa_v^2}}+e^{2z\sqrt{k^2+\kappa_v^2}}\right]\nonumber
\eea where $\Delta(x)$ is given in Eq.~(\ref{defDelta}). The
variational parameter of the Green's function is the variational
inverse screening length $\kappa_v$ which is taken uniform
(generalized Onsager-Samaras approximation, see
~\cite{yarosch,hatlo}). A more complicated approach has been
previously developed in Ref.~\cite{hatlo} where the authors
introduced a piecewise form for the variational screening length,
i.e. $\kappa(z)=\kappa_v$ over a layer of size $h$ and
$\kappa_v=\kappa_b$ in the middle of the pore. Although this choice
is more general than ours, the minimization procedure with respect
to $\kappa_v$ is significantly longer than in our case and the
variational equation is much more complicated. Consequently, this
piecewise approach is not very practical when one wishes to study a
charged membrane where the external field created by the surface
charge considerably complicates the technical task (see
Section~\ref{chargedpore}). We show that the simple variational
choice adopted here captures the essential physics with less
computational effort.

As in Eq.~(\ref{CHPO}), the integral on the rhs. of
Eq.~(\ref{GreenConf}) takes into account both image charge and
solvation effects due to the two interfaces, whereas the first term
is the Debye result for the difference between the bulk and a
hypothetic bulk of inverse screening length $\kappa_v$. We should
emphasize that, in the present case, the spatial integrations in
Eqs.~(\ref{F12})-(\ref{FvarIV}) run over the confined space, that is
from $z=0$ to $z=d$. By substituting the solution
Eq.~(\ref{GreenConf}) into Eqs.~(\ref{FvarIII})-(\ref{Fc}) and
performing the integration over $z$, one finds~\cite{hatlo_review}
\bea\label{FGII}
&&\frac{F_2+F_3}{S}=\frac{d\kappa_v^3}{24\pi}+\frac{\Delta\kappa_v^2}{16\pi}\\
&&+\frac{\kappa_v^2}{4\pi}\int_1^\infty\mathrm{d}xx\ln\left[1-\bar\Delta^2(x)
e^{-2\kappa_vdx}\right]\nonumber\\
&&+\frac{\kappa_v^2}{8\pi}\int_1^\infty\mathrm{d}x\frac{\left(\bar\Delta(x)-\bar\Delta^3(x)\right)/x
-2\kappa_vd\bar\Delta^2(x)}{e^{2d\kappa_vx}-\bar\Delta^2(x)}\nonumber
\eea where we have defined
$\bar\Delta(x)=\Delta\left(\sqrt{x^2-1}\right)$.

The limiting case $\epsilon=0$ allows for closed-form expressions.
This limit is a good approximation for describing biological and
artificial pores characterized by an external dielectric constant
much lower than the internal one. In the following part of the work,
we will deal most of the time with the special case $\epsilon=0$,
unless stated otherwise. In this limit, Eq.~(\ref{FGII}) simplifies
to \bea\label{FG}
\frac{F_2+F_3}{S}&=&\frac{\kappa_v^3d}{24\pi}+\frac{\kappa_v^2}{16\pi}\left[1+2\ln\left(1-e^{-2d\kappa_v}\right)\right]\nonumber\\
&-&\frac{\kappa_v}{8\pi d}\mathrm{Li}_2\left(e^{-2d\kappa_v}\right)
-\frac{\mathrm{Li}_3\left(e^{-2d\kappa_v}\right)}{16\pi d^2} \eea
where $\mathrm{Li}_n(x)$ stands for the polylogarithm function and
$\xi(x)$ the Riemann zeta function (see Appendix~\ref{appendix}).
Within the same limit ($\epsilon=0$), $\Delta(x)=1$ and we obtain an
analytical expression for the Green's function Eq.~(\ref{GreenConf})
\bea\label{PMFdble}
w_0(z)&=& (\kappa_b-\kappa_v)\ell_B-\frac{\ell_B}{d}\ln\left(1-e^{-2d\kappa_v}\right)\nonumber\\
&+&\frac{\ell_B}{2d}\left[\beta\left(e^{-2d\kappa_v};1-\frac{z}{d},0\right)\right.\nonumber\\
&&
+\left.\frac{d}{z}e^{-2\kappa_vz}\hspace{0.5mm}_2\mathrm{F}_1\left(1,\frac{z}{d},1+\frac{z}{d},e^{-2d\kappa_v}\right)\right]
\eea where $\beta(x;y,z)$ is the incomplete Beta function and
$_2\mathrm{F}_1(a,b;c;d)$ the hypergeometric series. The definitions
of these special functions are given in Appendix~\ref{appendix}. At
this step, the PMF thus depends on three adimensional parameters,
namely $d\kappa_v$, $d\kappa_b$, and $d/\ell_B$.

For the system with a single interface, the ion fugacity $\lambda_i$
was fixed by the bulk density. In the present case where the
confined system is in contact with an external reservoir,
$\lambda_i$ is fixed by chemical equilibrium: \be
\lambda_i=\lambda_{i,b}=\rho_{i,b}
e^{-\frac{q_i^2}2\kappa_{i,b}\ell_B},\label{IPE} \ee where
$\kappa_b$ and $\lambda_{i,b}$ are respectively the inverse Debye
screening length and the fugacity in the bulk reservoir [see
Eq.~(\ref{DH})]. Once this constraint is taken into account, the
last term of electrostatic part of the variational grand potential
Eq.~(\ref{FvarIII}) can be written as $-\sum_i\rho_{i,b}\int_0^d
\mathrm{d}z \;e^{-\frac{q_i^2}{2}w(z)-q_i\phi_0(z)}$.

Eq. (\ref{eqvarI}) then becomes for a symmetric $q:q$ electrolyte:
\be \frac{\partial^2 \tilde\phi_0}{\partial z^2}-\kappa_b^2
e^{-\frac{q^2}2w(z)}\sinh\tilde\phi_0 =-4\pi q
\ell_B\sigma_s\left[\delta(z)+\delta(z-d)\right]\label{eqvarP} \ee
The optimization of $F_v=F_1+F_2+F_3$ given by Eq. (\ref{FvarIII})
and (\ref{FG}) with respect to the inverse trial screening length
$\kappa_v$ leads to the following variational equation for
$\kappa_v$: \bea
(d\kappa_v)^2&+&d\kappa_v\tanh(d\kappa_v)=(d\kappa_b)^2 \int_0^1 \mathrm{d}x \,e^{-\frac{q^2}2 w_0(xd)}\nonumber\\
&\times& \cosh
[\tilde\phi_0(xd)]\left\{1+\frac{\cosh\left[(2x-1)d\kappa_v\right]}{\cosh(d\kappa_v)}\right\}.
\label{eqvarKap} \eea Within the particular choice that fixed the
functional form of the $\kappa_v$ dependent Green's function
Eq.~(\ref{PMFdble}), the two coupled equations~(\ref{eqvarP})
and~(\ref{eqvarKap}) are the most general variational equations. In
the following, we first consider the case of neutral pores and then
the more general case of charged pores.

\subsection{Neutral pore, symmetric electrolyte}
\label{neutral_pore}

\begin{figure}[t]
\includegraphics[width=1\linewidth]{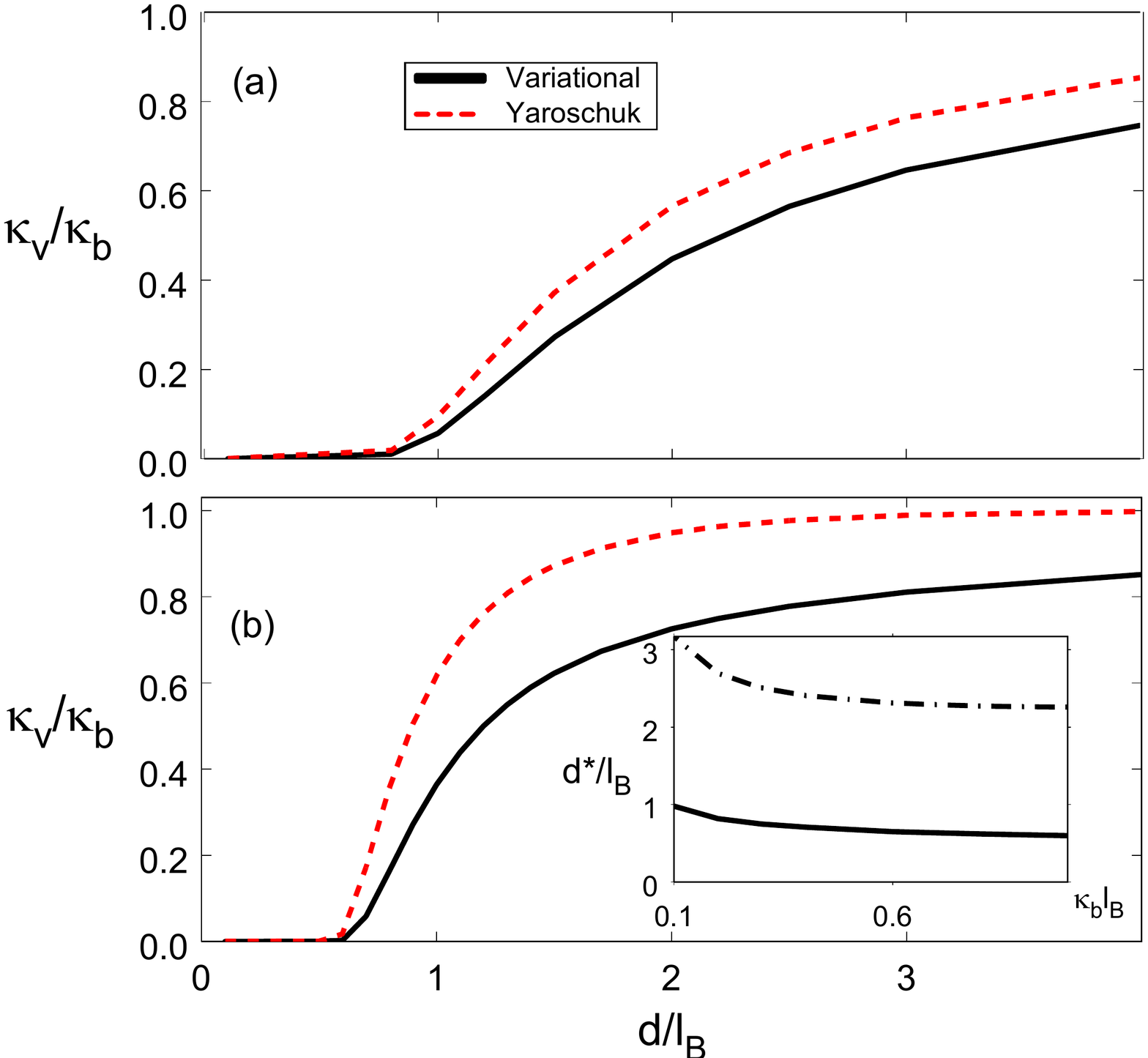}
\caption{(color online) Inverse screening length inside the neutral
membrane (monovalent ions) normalized by $\kappa_b$ vs. the pore
size $d/\ell_B$ for $\epsilon=0$ and (a) $\kappa_b\ell_B=0.1$
($\rho_b=1.926$ mM), (b) $\kappa_b\ell_B=1$ ($\rho_b=0.1926$ M).
Dashed lines correspond to the mid-point approximation,
Eq.~(\ref{vareqYarII}). The inset shows the characteristic pore size
corresponding to total ionic exclusion as a function of the inverse
bulk screening length. The bottom curve corresponds to monovalent
ions and the top curve to divalent ions.} \label{kvVShI}
\end{figure}

In the case of  a symmetric $q:q$ electrolyte and a neutral
membrane, $\sigma_s=0$, the solution of Eq.~(\ref{eqvarP}) is
naturally $\phi_0=0$. The variational parameter $\kappa_v$ is
solution of Eq.~(\ref{eqvarKap}) with $\phi_0=0$ and $w(z)=w_0(z)$
given by Eq.~(\ref{PMFdble}) when $\epsilon=0$, which can be written
as $d\kappa_v=f(d\kappa_b,\ell_B/d)$. Let us note that
Eq.~(\ref{eqvarKap}) can be solved with the Mathematica software  in
a fraction of a second.

Within the Debye-H\"uckel closure approach, Yaroshchuk (see Eq.~(59)
of Ref.~\cite{yarosch}) obtains a self-consistent approximation for
constant $\kappa_v$ by replacing the exponential term of
Eq.~(\ref{VarNetzG}) with its average value in the pore: \be
\kappa_v^2= \kappa_b^2 \int_0^1 \mathrm{d}x\,e^{-\frac{q^2}2 w(xd)},
\label{vareqYarI} \ee which should be compared with
Eq.~(\ref{eqvarKap}) with $\phi_0=0$. In order to simplify the
numerical task, Yaroshchuk introduces a further approximation in
which he replaces the potential $w(z)$ inside the depletion term of
Eq.~(\ref{vareqYarI}) by its value in the middle of the pore,
$w(d/2)$. Then Eq.~(\ref{vareqYarI}) takes the simpler form \be
\kappa_v^2=\kappa_b^2 e^{-\frac{q^2}{2}w(d/2)}. \label{vareqYarII}
\ee The self-consistent midpoint approximation is frequently used in
nanofiltration theories~\cite{yarosch,Szymczyk,YarII}. For
$\epsilon=0$, the mid-point potential has the simple form
$w(d/2)=(\kappa_b-\kappa_v)\ell_B-2\ell_B\ln(1-e^{-\kappa_v d})/d$.
This approach is compared with the full variational treatment in
Fig.~\ref{kvVShI}  where the adimensional inverse screening length
in the pore $\kappa_v/\kappa_b$ is plotted as a function of the pore
size $d$. We first note that as $d$ decreases below a critical value
$d^*$, the pore is empty of salt and $\kappa_v=0$. The inset of Fig.
\ref{kvVShI} shows $d^*$ versus the inverse bulk screening length.
Searching for $d$ such that $\kappa_v=0$ in Eq. (\ref{eqvarKap})
leads to the same equation as Eq.~(\ref{vareqYarI}), thus the value
of $d^*$ is identical within both approaches. However,
Fig.~\ref{kvVShI} shows that the mid-point approximation,
Eq.~(\ref{vareqYarII}), overestimates the internal salt
concentration as well as the abruptness of the crossover to an
ion-free regime  for decreasing pore size. Indeed, this
approximation is equivalent to neglecting the strong ion exclusion
close to the pore surfaces (which is larger than in the middle of
the pore). A similar behavior was also observed in Fig.~6 of
Ref.~\cite{hatlo} for the screening length in the neighborhood of
the dielectric interface.

\begin{figure}[t]
(a)\includegraphics[width=.8\linewidth]{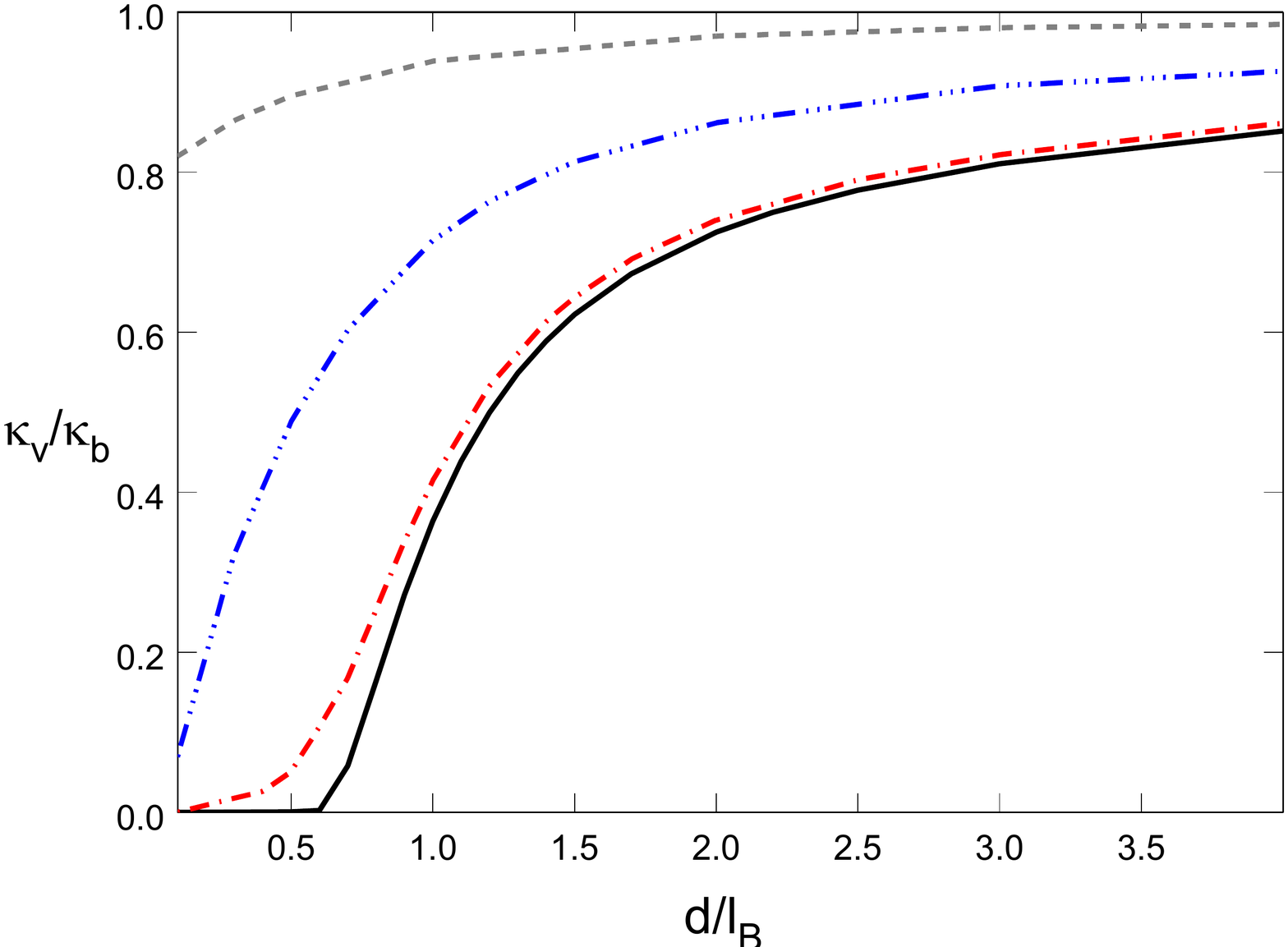}
(b)\includegraphics[width=.9\linewidth]{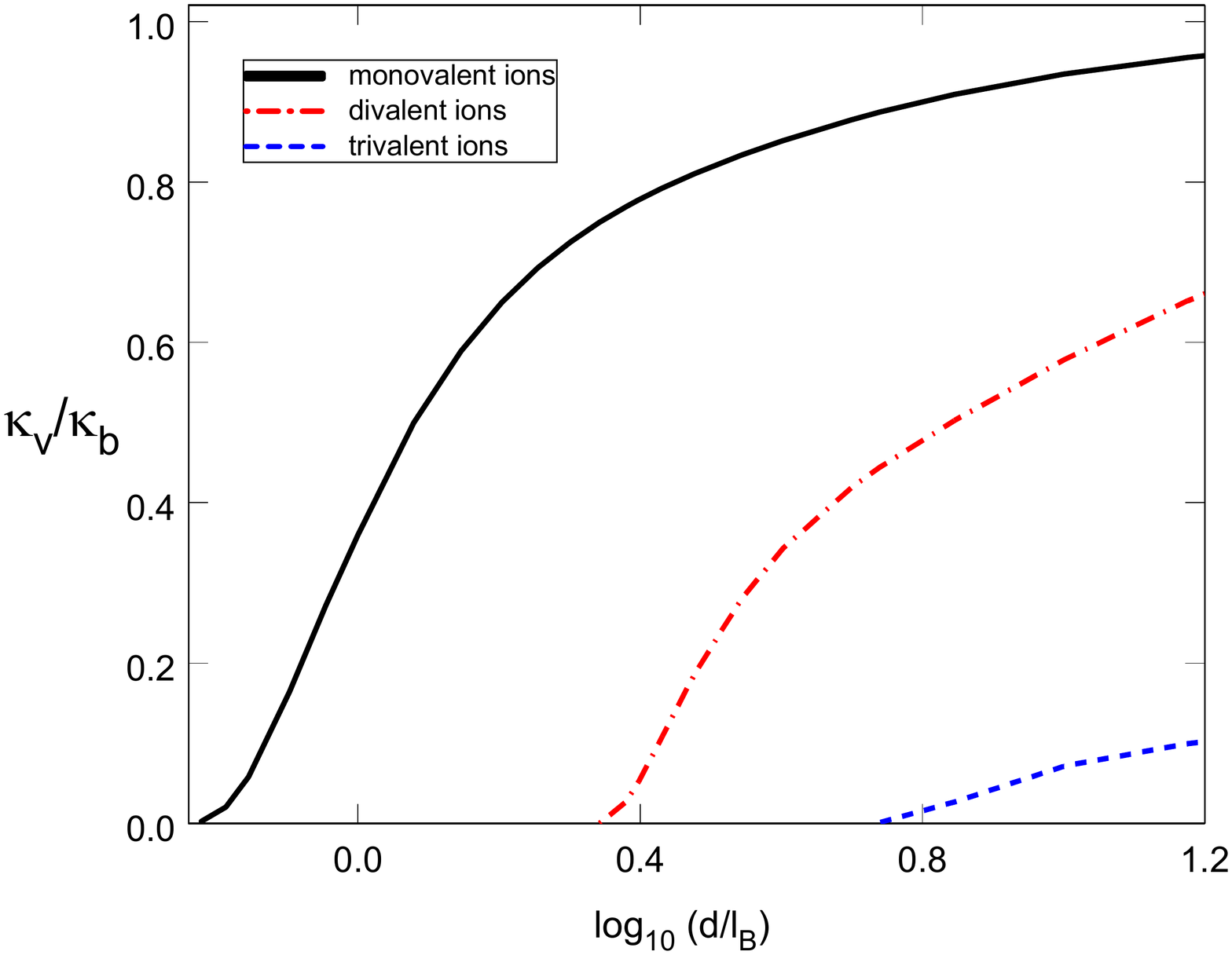}
\caption{(color online) Inverse screening length inside the membrane
vs. the pore size $d/\ell_B$ ($\epsilon_w=78$, $\kappa_b\ell_B=1$).
(a) From bottom to top: $\epsilon=0$ ($\Delta=1$), $\epsilon=3.2$
($\Delta=0.92$), $\epsilon=39$ ($\Delta=1/3$), and  $\epsilon=78$
($\Delta=0$). (b) Log-linear plot for monovalent, divalent and
trivalent ions, from left to right ($\epsilon=0$).}
\label{kvVShepsDIFF}
\end{figure}

The effect of the dielectric discontinuity is illustrated in
Fig.~\ref{kvVShepsDIFF}(a) where the inverse internal screening
length is compared for $\epsilon$ between 0 and $\epsilon_w=78$
where the image-charge repulsion is absent and the solvation effect
is solely responsible for ion repulsion. First of all, one observes
that the total exclusion of ions in small pores is specific to the
case $\epsilon=0$. Moreover, in the solvation only case, the inverse
screening length inside the pore only slightly deviates from the
bulk value, $0.8\leq\kappa_v/\kappa_b\leq1$. This clearly indicates
that, within the point-like ion model considered in this work, the
image-charge interaction brings the main contribution to
salt-rejection from neutral membranes. Roughly speaking, the
image-charge and solvation effects come into play when the surface
of the ionic cloud of radius $\kappa_b^{-1}$ around a single ion
located at the pore center touches the pore wall, i.e. for
$\kappa_b^{-1}>d/2$. This simple picture fixes a characteristic
length $d_{ch}\simeq 2\kappa_b^{-1}$ below which the internal ion
density significantly deviates from the bulk value and ion-rejection
takes place. This can be verified for intermediate salt densities in
the bottom plot of Fig.~\ref{kvVShI} and the top plot of
Fig.~\ref{kvVShepsDIFF}.

Since image-charge effects are proportional to $q^2$, we illustrate
in Fig.~\ref{kvVShepsDIFF}(b) the effect of ion valency $q$. At pore
size $d\simeq 2.5\ell_B\simeq 1.8$~nm, where the inverse internal
screening length for monovalent ions is close to $80\hspace{1mm}\%$
of its saturation value $\kappa_b$, the exclusion of divalent ions
from the membrane is total. This effect driven by image interactions
is even much more pronounced for trivalent ions. Since the typical
pore size of nano-filtration membranes ranges between $0.5$ and
$2$~nm, we thus explain why ion valency can play a central role in
ion selectivity, even inside neutral pores.
\begin{figure}
\includegraphics[width=.9\linewidth]{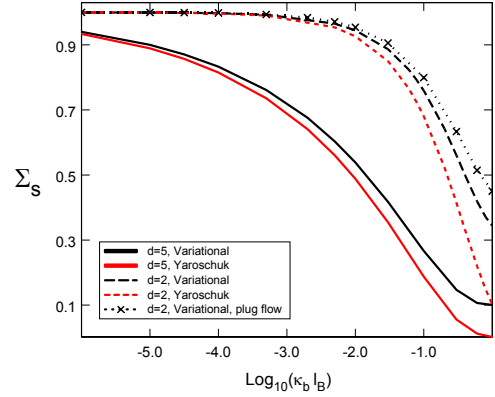}
\caption{(color online) Salt reflection coefficient (dimensionless) against the logarithm of the inverse
bulk screening length for $\epsilon=0$ and two pore sizes,
$d/\ell_B=2$ and 5 (red lines correspond to the mid-point
approximation, Eq.~(\ref{vareqYarII})).} \label{REJECT}
\end{figure}

The salt reflection coefficient, frequently used in membrane
transport theories to characterize the maximum salt-rejection
(obtained at high pressure) is related to the ratio of the net flux
of ions across the membrane to that of the solvent volume flux $J$
per unit transverse surface : \bea
\Sigma_s &\equiv& 1-\frac1{J\rho_b}\int_0^d  v_{||}(z) \rho(z) \mathrm{d}z \\
&=& 1-12\int_{1/2}^1 x(1-x)\;e^{-\frac{q^2}{2}w(xd)}
\mathrm{d}x\nonumber. \eea where we have used, in the second
equality, the Poiseuille velocity profile, $v_{||}(z)=\frac{6J}{d^3}
z(d-z)$ in the pore and the PMF given by Eq.~(\ref{defPMF}). It
depends only on the parameters $\kappa_b\ell_B$ and $d/\ell_B$. In
certain nanopores with hydrophobic surfaces, the solvent flux may
considerably deviate from the Poiseuille profile (see~\cite{PNAS}).
In this case, the velocity profile is flat, $v_{||}(z)=\frac{J}{d}$.
We emphasize that since the velocity profile is normalized in both
cases, the mid-point approximation is unable to distinguish between
a Poiseuille and a plug flow velocity profile. Fig.~\ref{REJECT}
displays $\Sigma_s$  as a function of the inverse bulk screening
length for two pore sizes $d=2\ell_B$ and $d=5\ell_B$. As seen by
Yaroshchuk, decreasing the pore size shifts the curves to higher
bulk concentration and thus increases the range of bulk
concentration where nearly total salt rejection occurs. However,
quantitatively, the difference between the variational and mid-point
approaches becomes significant at high bulk concentrations and this
difference is accentuated in the case of plug-flow (for which $\Sigma_s$ is higher when compared
to the Poiseuille case because the flow velocity no longer vanishes at the pore wall where the salt exclusion
is strongest). This deviation is again due to the midpoint approximation of Eq.~(\ref{vareqYarII})
in which the image interactions are underestimated. However since
the velocity profile vanishes at the solid surface for the
Poiseuille flow, the deficiencies of the mid-point approximation are
less visible in $\Sigma_s$ than in $\kappa_v$ in this case.

Finally, we compute the disjoining pressure within our variational
approach. We compare in Appendix \ref{appendixPR} the result  with
that of the more involved variational scheme presented in
Ref.~\cite{hatlo} and show that one gets a very similar behaviour,
revealing that the simpler variational method is able to capture the
essential physics of the slit pore.

As stressed above, the main benefit obtained from the simpler
approach proposed in this work is that the minimization procedure is
much less time consuming. This point becomes crucial when
considering the fixed charge of the membrane, which is thoroughly
studied in the next section.

\subsection{Charged pore, symmetric electrolyte}
\label{chargedpore}

In this section, we apply the variational approach to a slit-like
pore of surface charge $\sigma_s<0$.  In the following, we will
solve Eqs.  (\ref{eqvarP}) and (\ref{eqvarKap}) numerically in order
to test, as in the case of a single charged surface, the validity of
restricted trial forms for $\phi_0(z)$. We define the partition
coefficients in the pore for counterions and coions, $k_+$ and
$k_-$, as \be\label{AvDen}
k_{\pm}\equiv\frac{\rho_{\pm}}{\rho_b}=\int_0^d\frac{\mathrm{d}z}{d}
e^{-\Phi_{\pm}(z)}. \ee where $\Phi_{\pm}(z)$ is given by
Eq.~(\ref{PMF2}).

\subsubsection{\underline{Effective Donnan Potential}}

When one considers a charged nanopore, because of its small size,
gradients of the potential $\phi_0$ can be neglected as a first
approximation. We thus assume a constant potential $\bar{\phi}_0$.
The so-called effective \textit{Donnan potential} $\bar{\phi}_0$
introduced by Yaroshchuk \cite{yarosch}  will be fixed by the
variational principle. By differentiating the grand potential
Eq.~(\ref{FvarIII}) with respect to $\bar{\phi}_0$ (or equivalently
integrating Eq.~(\ref{eqvarP}) from $z=0$ to $z=d$ with
$\nabla\bar{\phi}_0=0$), we find \be\label{EN}
2\left|\sigma_s\right|=-2q\rho_b\sinh(q\bar{\phi}_0)\int_0^d
\mathrm{d}z \,e^{-\frac{q^2}2w(z)} \ee which is simply the
electroneutrality relation in the pore, taken in charge by the
electrostatic potential $\bar{\phi}_0$. By defining \bea\label{GMA}
\Gamma&=&\int_0^1 \mathrm{d}x \exp[-q^2\ell_B\bar{w}(xd)/(2d)]\\
&=&\int_0^1 \mathrm{d}x \exp[-\Xi \bar{w}(x\tilde d)/(2\tilde d)],
\eea where $\bar{w}(x)\equiv w(x)d/\ell_B$, we have
$k_\pm=\Gamma\exp(\mp q\bar{\phi}_0)$ and Eq.~(\ref{EN}) can be
rewritten as \be\label{partdif} k_+-k_-=2\frac{|\sigma_s|}{q\rho_b
d}=\frac{X_m}{q\rho_b}= \frac8{\kappa_b^2d\ell_G}=
\frac8{\tilde{\kappa}_b^2\tilde{d}} \ee where the second equality
contains the Gouy-Chapman length $\ell_G$ and the quantity
$X_m=2|\sigma_s|/d$, frequently used in nanofiltration theories,
corresponds to the volume charge density of the membrane. Hence, the
partition coefficient of the charge, $k_+-k_-$, does not depend on
$\Xi$, i.e. charge image and solvation forces.
\begin{figure}[t]
\includegraphics[width=0.95\linewidth]{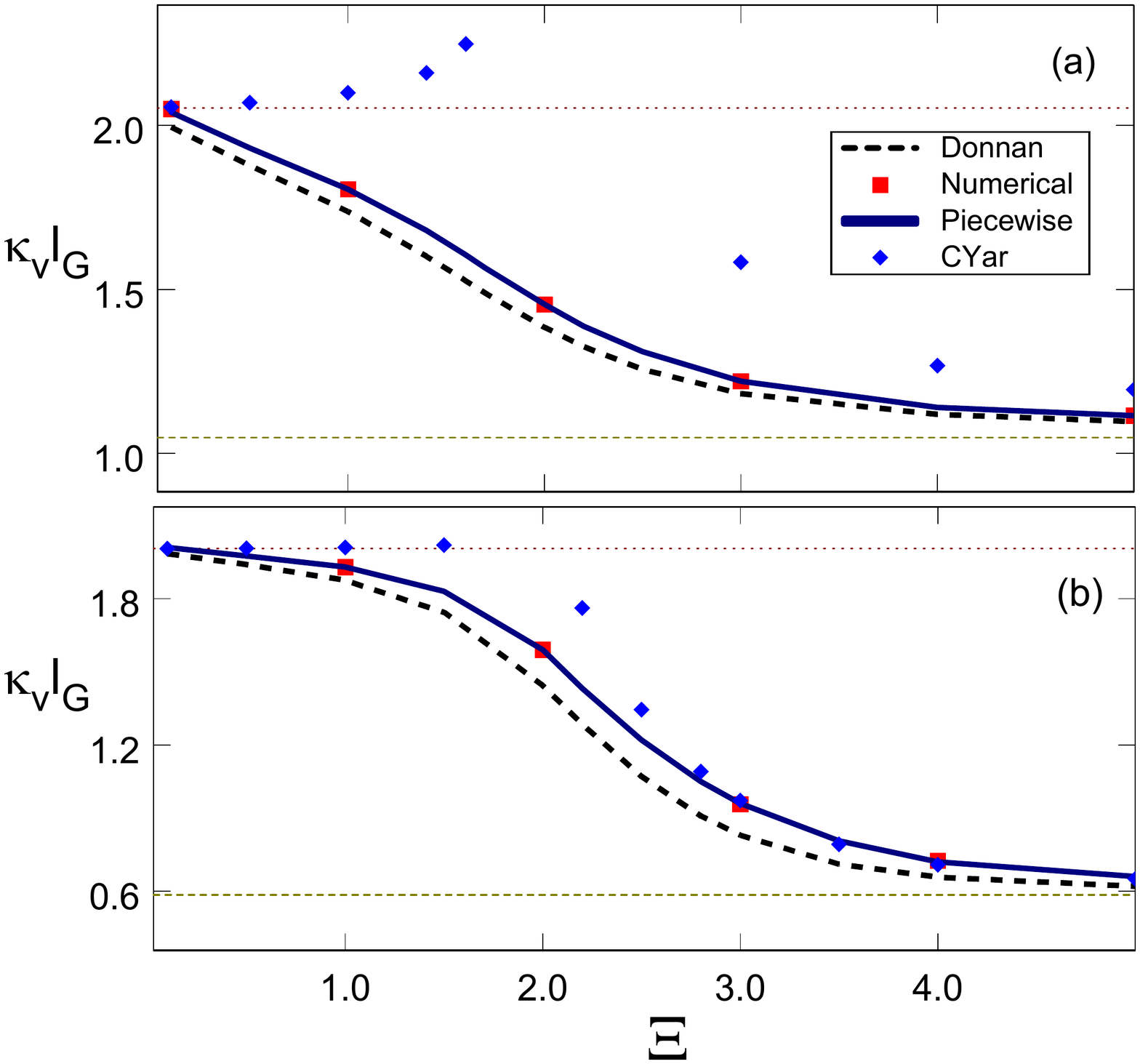}
\caption{(color online) Inverse internal screening length $\kappa_v$
against $\Xi$ for $\kappa_b\ell_G=2$, $\epsilon=0$ and (a)
$d=3\ell_G$ and (b) $d=10\ell_G$. Comparison of various
approximations: Yaroshchuk, Eq.~(\ref{CYarI}) (diamonds),
variational Donnan potential (dashed line), piecewise solutions
(solid line), and numerical results (squares). Horizontal lines
corresponds to the WC limit, Eq.~(\ref{asymWC}) (top), and SC limit,
Eq.~(\ref{asymSCII}) (bottom) .} \label{KPvsKSI}
\end{figure}
By using Eq.~(\ref{EN}) in order to eliminate the potential
$\bar{\phi}_0$ from Eq.~(\ref{AvDen}), one can rewrite the partition
coefficients in the form \be\label{partcoef} k_{\pm}= \Gamma e^{\mp
q\bar\phi_0} = \sqrt{\Gamma^2+\left(\frac4{\tilde\kappa_b^2\tilde
d}\right)^2} \pm\frac4{\tilde\kappa_b^2\tilde d} \ee By substituting
into Eq.~(\ref{eqvarKap}) the analytical expression for
$\bar{\phi}_0$ obtained from Eq.~(\ref{EN}) (or
Eq.~(\ref{partcoef})), one obtains a single variational equation for
$\kappa_v$ to be solved numerically, \bea\label{DNV}
&&(\tilde{d}\tilde{\kappa}_v)^2+\tilde{d}\tilde{\kappa}_v\tanh(\tilde{d}\tilde{\kappa}_v)=
(\tilde{d}\tilde{\kappa_b})^2\sqrt{\Gamma^2+\left(\frac4{\tilde\kappa_b^2 \tilde d}\right)^2}\nonumber\\
&\times&\left\{1+\int_0^1 \mathrm{d}x \,e^{-\frac{\Xi}{2}
\tilde{w}(x\tilde{d})}\frac{\cosh\left[(2x-1)\tilde{d}\tilde{\kappa}_v\right]}{\Gamma\cosh(\tilde{d}\tilde{\kappa}_v)}\right\}.
\eea The numerical solution of Eq.~(\ref{DNV}) is plotted in
Fig.~\ref{KPvsKSI} as a function of the coupling parameter $\Xi$. We
see that as we move from the WC limit to the SC one by increasing
$\Xi$, the pore evolves from a high to a low salt regime. This quite
rapid crossover, which results from the exclusion of ions from the
membrane, is mainly due to repulsive image-charge and solvation
forces controlled by $\Gamma$ whose effects increase with increasing
$\Xi$.

\begin{figure}
\includegraphics[width=1.1\linewidth]{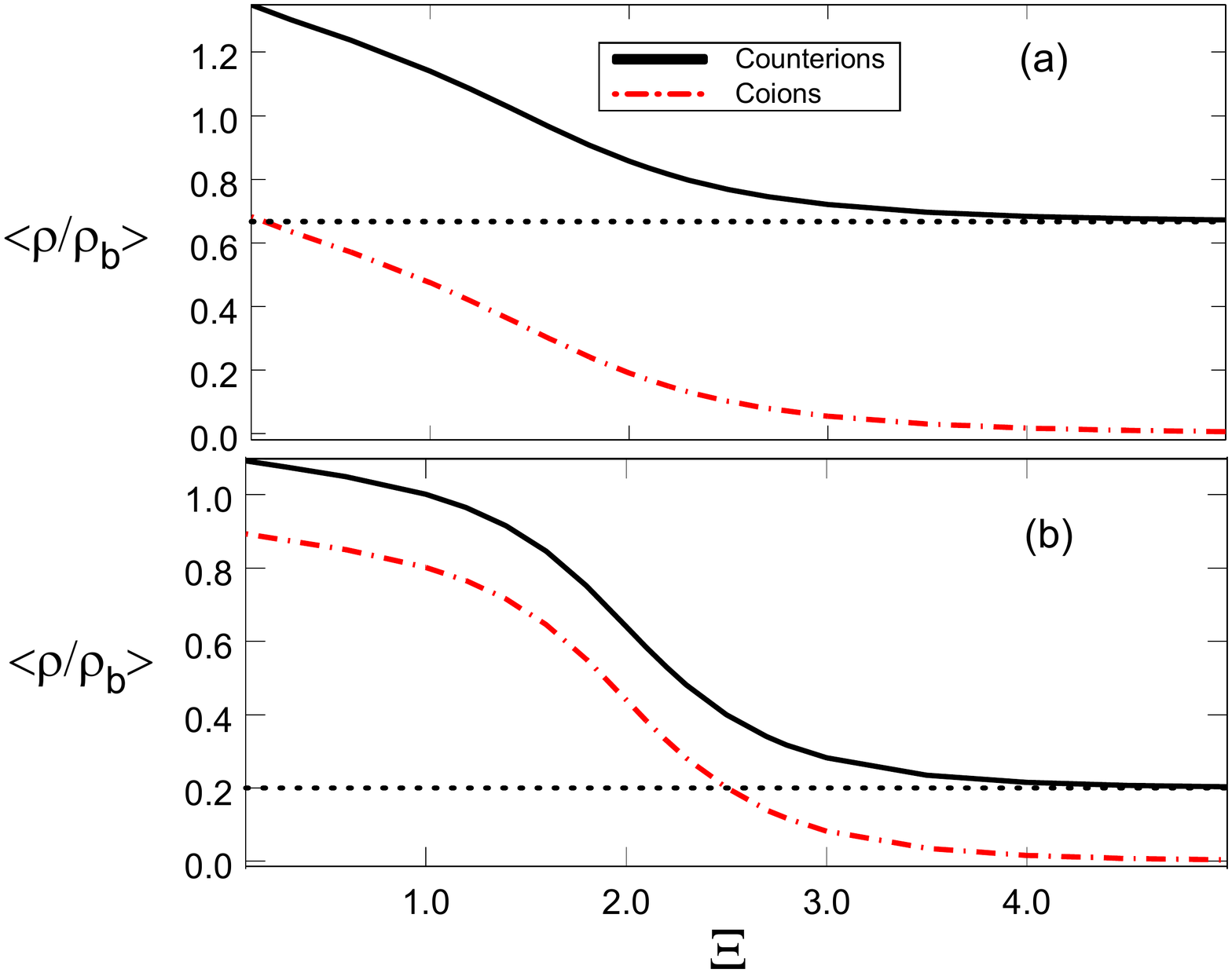}
\caption{(color online) Ionic partition coefficients, $k_\pm$, vs.
$\Xi$ for $\kappa_b\ell_G=2$, $\epsilon=0$, and (a) $d=3\ell_G$ and
(b) $d=10\ell_G$. The horizontal line corresponds to the SC limit
for counterions. As explained in the text,we note that
$k_+-k_-=8/(\kappa_b^2d\ell_G)$.} \label{DenvsKSI}
\end{figure}

In Fig.~\ref{DenvsKSI} are plotted the partition coefficients of
counterions and coions, Eq.~(\ref{AvDen}), as a function of $\Xi$.
Here again, $k_\pm$ decreases with increasing $\Xi$. Moreover, we
clearly see that the rejection of coions from the membrane becomes
total for $\Xi>4$. In other words, even for intermediate coupling
parameter values, we are in a counterion-only state. This is
obviously related to the electrical repulsion of coions by the
charged surface.

In the asymptotic WC limit ($\Xi\to0$), $\Gamma=1$ and we find the
classical Donnan results in mean-field where
$k_-=k^{-1}_+=e^{q\bar{\phi}_0}$ with $q\bar{\phi}_0={\rm arcsinh}
[4/(\tilde \kappa_b^2 \tilde d)]$. The variational
equations~(\ref{DNV}) and~(\ref{partcoef}) reduce to \bea
\kappa_v^2 &=& \kappa_b^2\sqrt{1+\left(\frac4{\tilde \kappa_b^2 \tilde d}\right)^2}\label{asymWC}\\
k_{\pm} &=& \sqrt{1+\left(\frac4{\tilde\kappa_b^2\tilde d}\right)^2}
\pm\frac4{\tilde\kappa_b^2\tilde d}\label{AvDenWC} \eea Quite
interestingly, the relation Eq.~(\ref{asymWC}) shows that, even in
the mean-field limit, due to the ion charge imbalance created by the
pore surface charge, the inverse screening length is larger than the
Debye-H\"uckel value $\kappa_b$.  In the case of small pores or
strongly charged pores or at low values of the bulk ionic strength,
i.e. $\kappa_b^2\ell_G d\ll 1$ or $d\rho_b\ll |\sigma_s|/q$,  we
find $\kappa_v\simeq2/\sqrt{\ell_G d}$ and $\rho_-=0$ and
$\rho_{+}=2|\sigma_s|/(dq)$. We thus find the classical
Poisson-Boltzmann result for counterions only~\cite{netz_WSC}. The
counterion-only case is also called \textit{good coion exclusion}
limit (GCE), a notion introduced in the context of nanofiltration
theories~\cite{yarosch,lefebvre,Yar}.  Hence, in this limit the
quantity of counterions in the membrane is independent of the bulk
density and depends only on the pore size $d$ and the surface charge
density $\sigma_s$. In the case of a membrane of size $d\simeq1$~nm
and fixed surface charge $\sigma_s\simeq 0.03\;\mbox{nm}^{-2}$, this
limit can be reached with an electrolyte of bulk concentration
$\rho_b\simeq50$~$\mbox{mM}$. In the opposite limit
$\kappa_b^2\ell_G d\gg 1$, one finds $\kappa_v\simeq\kappa_b$ and
$\rho_{\pm}=\rho_b$.

In the SC limit $\Xi\to\infty$, $\Gamma=0$ and Eq.~(\ref{DNV})
simplifies to \be (\tilde d\tilde \kappa_v)^2+\tilde d\tilde
\kappa_v\,\tanh(\tilde d\tilde\kappa_v)= 4\tilde d
[1+\mbox{sech}(\tilde d\tilde\kappa_v)]. \label{asymSCI} \ee For
$d>\ell_G$ ($\tilde d>1$), the solution of Eq.~(\ref{asymSCI})
yields with a high accuracy \be \tilde
\kappa_v\simeq\frac{\sqrt{1+16\tilde d}-1}{2\tilde d}.
\label{asymSCII} \ee The partition coefficients simplify to $k_-=0$
and $k_{+}=8/(\tilde d\tilde \kappa_b^2) =2|\sigma_s|/(dq\rho_b)$
and we find the counterion only case (or GCE limit) without image
charge forces discussed by Netz~\cite{netz_WSC}. Partition
coefficients  in the SC limit and variational inverse screening
length in both limits, Eqs.~(\ref{asymWC}) and~(\ref{asymSCII}), are
illustrated in Figs.~\ref{KPvsKSI} and~\ref{DenvsKSI} by dotted
reference lines. Consequently, one reaches for $\Xi=0$ the GCE limit
exclusively for low salt density or small pore size, while the SC
limit leads to GCE for arbitrary bulk density. It is also important
to note that although the pore-averaged densities of ions are the
same in the GCE limit of WC and SC regimes, the density profiles are
different since when one moves away from the pore center, the
counterion densities close to the interface increase in the WC limit
due to the surface charge  attraction and decrease in the SC limit
due to the image charge repulsion.

It is interesting to compare this variational approach to the
approximate mid-point approach of Yaroshchuk~\cite{yarosch}. For
charged membranes, he considers a constant potential and replaces
the exponential term of Eqs.~(\ref{VarNetzP}) and~(\ref{VarNetzG})
by its value in the middle of the pore. He obtains the following
self-consistent equations: \bea
\kappa^2&=&\kappa_b^2 e^{-\frac{q^2}{2}w(d/2)}\cosh\left(q\bar{\phi}_0\right)\label{CYarI}\\
2\left|\sigma_s\right|&=&-2qd\rho_b\sinh\left(q\bar{\phi}_0\right)\hspace{0.5mm}e^{-\frac{q^2}{2}w(d/2)}.\label{CYarII}
\eea The above set of equations are frequently used in
nanofiltration theories~\cite{yarosch,Szymczyk,YarII}. By combining
these equations in order to eliminate $\bar{\phi}_0$, one obtains an
approximate non-linear equation for $\kappa_v$ (approximation CYar
in Fig.~\ref{KPvsKSI}). In the limit of a high surface charge, the
non-linear equations~(\ref{CYarI})--(\ref{CYarII}) depend only on
the pore size $d$ and the surface charge density $\sigma_s$:
\be\label{asymp} \kappa^2\simeq\frac{8\pi
\ell_Bq|\sigma_s|}{d}=\frac{4}{\ell_G d}. \ee One can verify that in
the regime of strong surface charge, Eq.~(\ref{asymp}) is also
obtained from the asymptotic solution Eq.~(\ref{asymSCII}) since the
dependence of the PMF on $z$ is killed when $\Xi\to\infty$ and only
the mid-pore value contributes.
\begin{figure}
\includegraphics[width=1.1\linewidth]{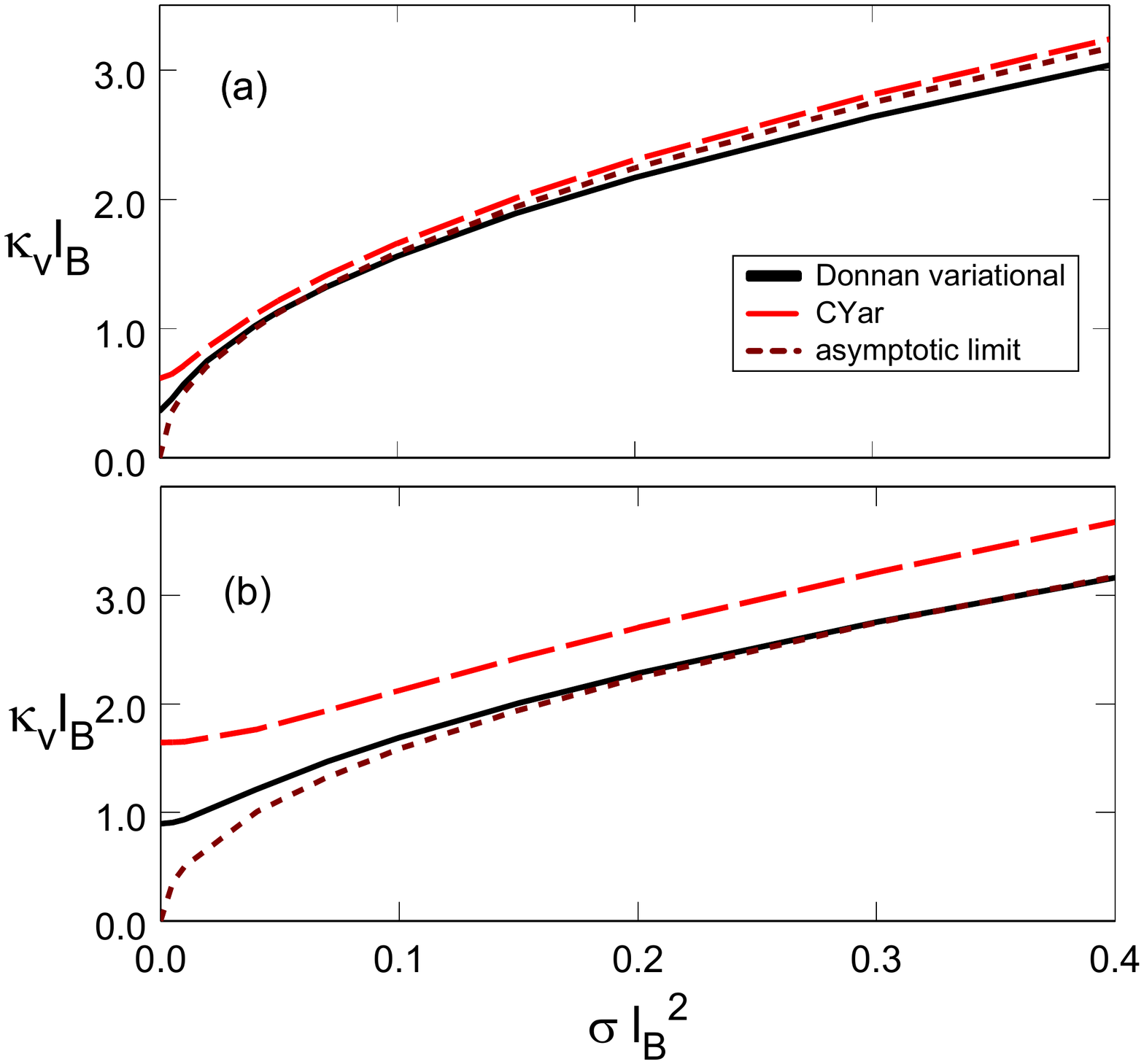}
\caption{(color online) Inverse internal screening length $\kappa_v$
against the reduced surface charge $\bar{\sigma}=\ell_B^2\sigma_s$
for $d=\ell_B$, $\epsilon=0$ and (a) $\kappa_b\ell_B=1$, (b)
$\kappa_b\ell_B=2$: constant variational Donnan approximation (solid
line), asymptotic result Eq.~(\ref{asymp}) (dotted line) and
Yaroshchuk approximation Eq.~(\ref{CYarI}) (dashed line).}
\label{KPvsSIG}
\end{figure}
The numerical solution of Eq.~(\ref{CYarI})-(\ref{CYarII}) is
illustrated as a function of $\Xi$  in Fig.~\ref{KPvsKSI}, and as a
function of the surface charge in Fig.~\ref{KPvsSIG}, together with
the asymptotic formula Eq.~(\ref{asymp}). For the parameter range
considered in Fig.~\ref{KPvsKSI}, the solution of Eq.~(\ref{CYarI})
strongly deviates from the result of the full variational
calculation. For $\Xi<2$, the mid-point approach follows an
incorrect trend with increasing $\Xi$. It is clearly seen that at
some values of the coupling parameter,
Eqs.~(\ref{CYarI})-(\ref{CYarII}) do not even present a numerical
solution. Using the relations $d/\ell_B=\tilde d/\Xi$ and
$\ell_B\kappa_b=\Xi\tilde{\kappa}_b$ for monovalent ions, one can
verify that the regime where the important deviations take place
corresponds to high ion concentrations. This is confirmed in
Fig.~\ref{KPvsSIG}: the error incurred by the approximate mid-point
solution of Yaroshchuk increases with the electrolyte concentration.
\\

\begin{figure}[t]
\includegraphics[width=.9\linewidth]{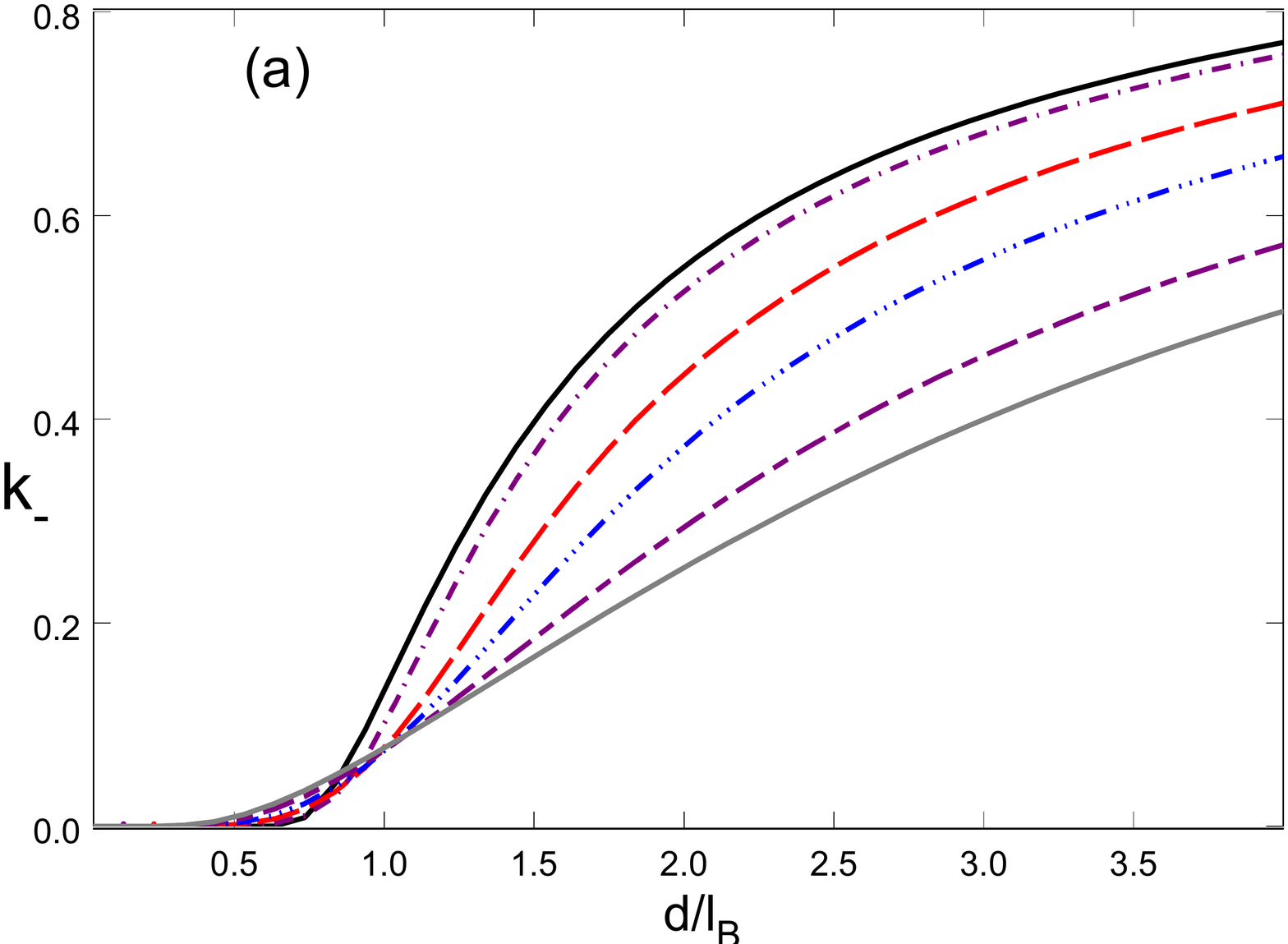}
\includegraphics[width=.9\linewidth]{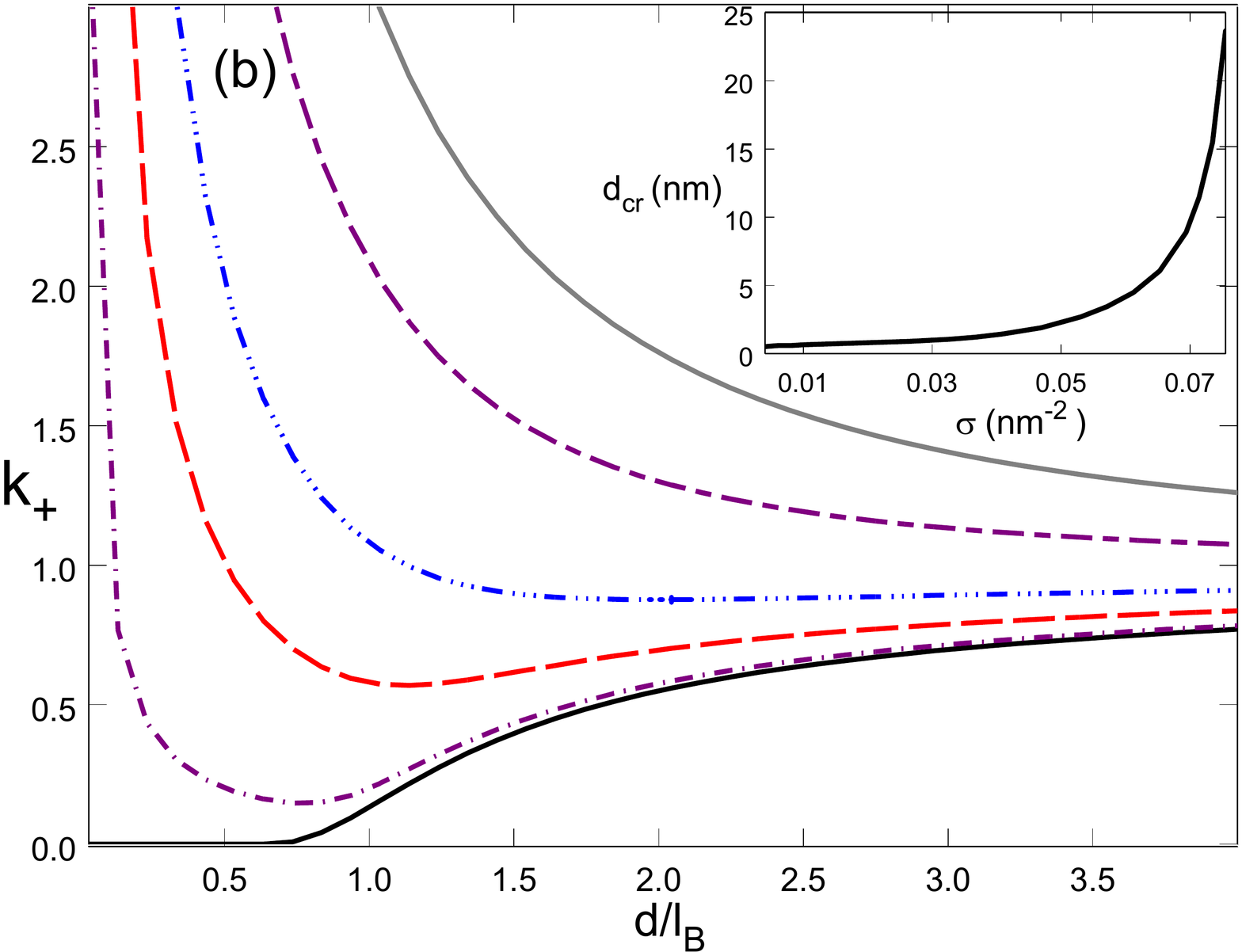}
\caption{(color online) Partition coefficient in the pore of coions
(a) and counterions (b) vs. the pore size $d/\ell_B$ for increasing
surface charge density,
$\sigma_s\ell_B^2=0,0.004,0.02,0.04,0.08,0.12$, from left to right,
and $\kappa_b\ell_B=1$. \textit{Inset}: Critical pore size $d_{\rm
cr}$ vs. the surface charge density $\sigma_s$ ($\epsilon=0$).}
\label{KPvshANDsig}
\end{figure}
In Section~\ref{neutral_pore} on neutral nanopores, it has been
underlined that, due to the image charge repulsion, the ionic
concentration inside the pore increases with the pore size $d$ (see
Fig.~\ref{kvVShepsDIFF}). In the present case of charged nanopores,
this result is modified: Eqs.~(\ref{asymWC}),~(\ref{asymSCII})
and~(\ref{asymp}) show that for strongly charged nanopores the
concentration of ions inside the pore decreases with $d$. Moreover,
the very high charge limit is a counterion-only state and
Eq.~(\ref{EN}) shows that, for a fixed surface charge density,
electroneutrality alone fixes the number of counterions, $N_-$, in a
layer of length $d$ joining both interfaces, and image charge
interactions play a little role. This is the reason why
$\kappa_v^2\propto\rho_-=N_-/(Sd)$ decreases for increasing $d$.

Hence, we expect an intermediate charge regime which interpolate
between image force counterion repulsion (case of neutral pores, see
Section~\ref{neutral_pore})  and counterion attraction by the fixed
surface charge. This is illustrated in Fig.~\ref{KPvshANDsig} where
the partition coefficients are plotted vs. $d$ for increasing
$\sigma_s$. As expected, coions are electrostatically pushed away by
the surface charge which adds to the repulsive image forces, leading
to a stronger coion exclusion than for neutral pores. The issue is
more subtle for counterions: obviously, increasing the surface
charge, $\sigma_s$, at constant pore size, $d$, increases $k_+$.
However, for small fixed $\sigma_s$, a regime where image charge and
direct electrostatic forces compete, $k_+$ is non-monotonic with
$d$.  Below a characteristic pore size, $d<d_{\rm cr}$, the
electrostatic attraction dominates over image charge repulsion and
due to the mechanism explained above, $k_+$ decreases for increasing
$d$. For $d>d_{\rm cr}$, the effect of the surface charge  weakens
and $k_+$ starts increasing with $d$. In this regime, the pore
behaves like a neutral system. The inset of Fig.~\ref{KPvshANDsig}
shows that $d_{\rm cr}$ increases when $\sigma_s$ increases. For
highly charged membranes $l_B^2\sigma_s\gg0.1$, there is no minimum
in $k_+(d)$, and the average counterion density inside the membrane
monotonically decreases towards the bulk value. Experimental values
for surface charges are $0\leq\sigma_s\leq0.5\;\mathrm{nm}^{-2}$ (or
$0\leq\ell_B^2\sigma_s\leq0.25$), which corresponds to physically
attainable values of $d_{\rm cr}$. The interplay between image
forces and direct electrostatic attraction is thus relevant to the
experimental situation.

The variational Donnan potential approximation is thus of great
interest since it yields physical insight into the exclusion
mechanism and allows a reduction of the computational complexity.
However, membranes and nanopores are often highly charged and
spatial variations of the electrostatic potential inside the pore
may play an important role. In the following we seek a piecewise
solution for $\phi_0(z)$.
\begin{figure}[t]
\includegraphics[width=1\linewidth]{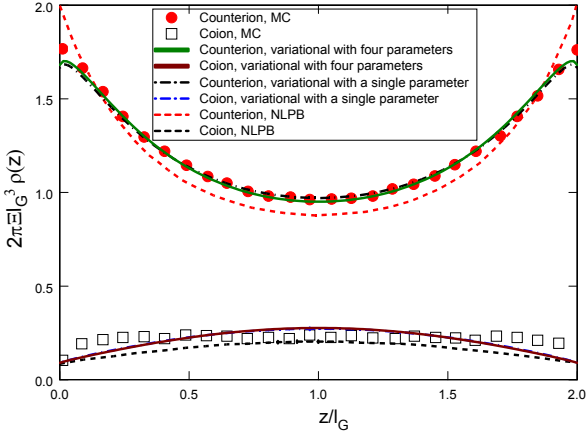}
\caption{(color online) Ion densities in the nanopore for
$\epsilon=\epsilon_W$, $\Xi=1$ and $h/\ell_G=2$. The continuous
lines correspond to the prediction of the variational method with
four parameters, the dashed-dotted line the variational solution
with a single parameter (see the text), the symbols are MC results
(Fig. 2 of~\cite{Li}) and the dashed lines denote the numerical
solution of the non-linear PB result.} \label{MCcompar}
\end{figure}
\subsubsection{\underline{Piecewise solution}}

The variational modified PB equation~(\ref{eqvarP}) for
$\tilde\phi_0$ shows that as one goes closer to the dielectric
interface, $w(z)$ increases and the screening experienced by the
potential $\tilde\phi_0$ gradually decreases because of ionic
exclusion. This non-perturbative effect which originates from the
strong charge-image repulsion inspires our choice for the
variational potential $\tilde\phi_0(z)$. We opt for a piecewise
solution as in Section~\ref{singleinter}: a salt-free solution in
the zone $0<z<h$ and the solution of the linearized PB equation for
$h<z<d/2$, with a charge renormalization parameter $\eta$ taking
into account non-linear effects. By inserting the boundary
conditions $\partial\tilde\phi_0/\partial z |_{z=0}=2\eta/\ell_G$
and $\partial\tilde\phi_0/\partial z|_{z=d/2}=0$ and imposing the
continuity of $\tilde\phi_0$ and its first derivative at $z=h$
[Eq.~(\ref{Cont})], the piecewise potential, solution of
Eq.~(\ref{eqvarP}) with $\kappa_b^2 \exp[-q^2w(z)/2]$ replaced by
$\kappa^2_\phi$, takes the form \be
 \tilde\phi_0(z)=\left\lbrace
 \begin{array}{ll}
 \bar{\varphi}-\frac{2\eta}{\ell_G}\left|z-\frac{d}2\right| & \mathrm{for}\quad 0<z\leq h,\\
\varphi-\frac{2\eta}{\ell_G\kappa_\phi} \frac{\cosh[\kappa_\phi(d/2-z)]}{\sinh[\kappa_\phi(d/2-h)]} & \mathrm{for}\quad h\leq z\leq d/2
 \end{array}\right.
 \label{resL2}
\ee
\begin{figure}[t]
\includegraphics[width=1\linewidth]{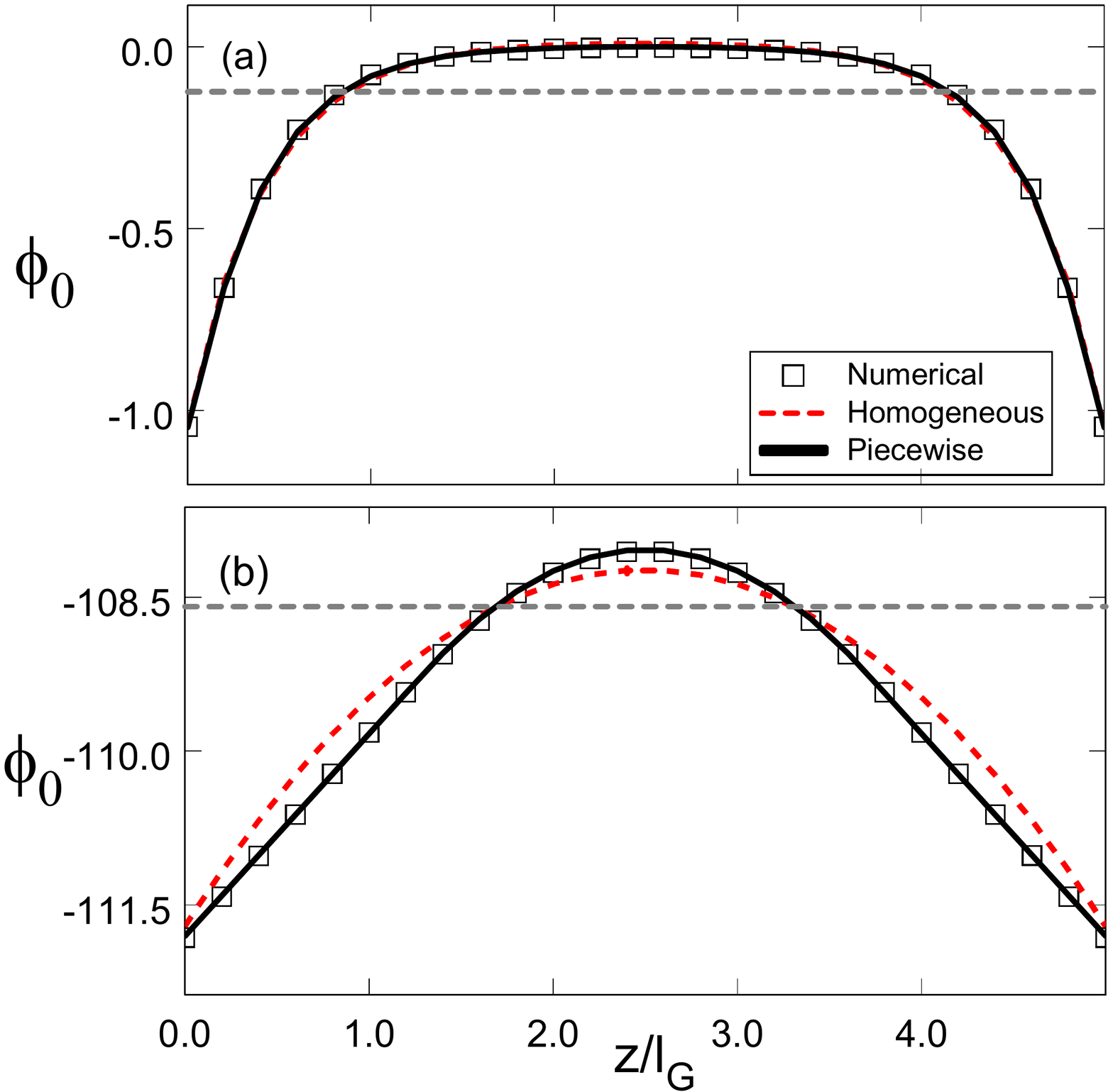}
\caption{(color online) Variational electrostatic potential (in
units of $k_BT$) in the nanopore. Comparison of the numerical
solution of Eq.~(\ref{eqvarP}) with the homogeneous ($h=0$) and
piecewise solution of Eq.~(\ref{resL2}) for $\ell_G\kappa_b=3$ and
(a) $\Xi=1$, (b) $\Xi=100$. The horizontal line is the Donnan
potential obtained from Eq.~(\ref{EN}) ($\epsilon=0$).}
\label{FieldSL}
\end{figure}
where \be
\bar{\varphi}=\varphi+\frac{2\eta}{\ell_G}\left(\frac{d}{2}-h\right)-
\frac{2\eta}{\ell_G\kappa_\phi}\coth\left[\kappa_\phi\left(\frac{d}2-h\right)\right]
\ee is imposed by continuity, and $\kappa_\phi$, $\varphi$, $h$ and
$\eta$ are the variational parameters. By injecting the piecewise
solution Eq.~(\ref{resL2}) into Eq.~(\ref{FvarIII}), we finally
obtain \bea
\frac{F_1}{S} &=& -\frac{2\left|\sigma_s\right|}{q} \left\{ \eta(\eta-2)\frac{h}{\ell_G} - \frac{\eta^2(d/2-h)}{2\ell_G\sinh^2[\kappa_\phi(d/2-h)]}\right. \nonumber\\
&+&\left. \frac{\eta(\eta-4)}{2\ell_G\kappa_\phi}\coth[\kappa_\phi(d/2-h)] +\varphi  \right\} \nonumber\\
&-&\frac{\kappa_b^2}{4\pi \ell_Bq^2}\int_0^d
\mathrm{d}z\;e^{-\frac{q^2}{2}w(z)}\cosh\tilde\phi_0(z). \eea The
solution to the variational problem is found by optimization of the
total grand potential $F=F_1+F_2$ with respect to  $\kappa_\phi$,
$\varphi$, $h$, $\eta$ and $\kappa_v$, where $F_2+F_3$ is given by
Eq.~(\ref{FG}) for a general value of $\epsilon$ and by
Eq.~(\ref{FG}) for $\epsilon=0$. This was easily carried out with
Mathematica software.

\textit{A posteriori}, we checked that two restricted forms for
$\phi_0$, homogeneous with $h=0$ and piecewise with $\varphi=0$,
were good variational choices. Fig.~\ref{MCcompar} compares the ion
densities obtained from the variational approach (with homogeneous
$\phi_0$) with the predictions of the MC simulations~\cite{Li} and
the NLPB equation for $\epsilon=\epsilon_W$, $\tilde d=2$ and
$\Xi=1$. Two variational choices are displayed in this figure,
namely, the homogeneous approach with four parameters
$\kappa_v=1.68,\kappa_\phi=1.36,\varphi=0.16, \eta=0.97$ and a
simpler choice with $\eta=1$, $\kappa_\phi=\kappa_v$ and two
variational parameters: $\kappa_v=1.69, \phi=-0.18$. In the latter
case, one can obtain an analytical solution for $\varphi$ and
injecting this solution into the free energy, one is left with a
single parameter $\kappa_v$ to be varied in order to find the
optimal solution. We notice that with both choices, the agreement
between the variational method and MC result is good. It is clearly
seen that the proposed approach can reproduce with a good
quantitative accuracy the reduced solvation induced ionic exclusion,
an effect absent at the mean-field level. Moreover, we verified that
with the single parameter choice, one can reproduce at the
mean-field variational level the ion density profiles obtained from
the numerical solution of the NLPB equation (dashed lines in
Fig.~\ref{MCcompar}) almost exactly. We finally note that the small
discrepancy between the predictions of the variational approach and
the MC results close to the interface may be due  to either
numerical errors in the simulation, or our use of the generalized
Onsager-Samaras approximation (our homogeneous choice for the
inverse effective screening length appearing in the Green's function
$v_0$ does not account for local enhancement or diminution of ionic
screening due to variations in local ionic density).

For $\epsilon=0$, the piecewise and homogeneous solutions are
compared with the full numerical solution of
Eqs.~(\ref{eqvarP})--(\ref{eqvarKap}) in Fig.~\ref{FieldSL} for
$\Xi=1$ and 100. First of all, one observes that  for $\Xi=1$, both
variational solutions match perfectly well with the numerical
solutions. For $\Xi=100$, the piecewise solution matches also
perfectly well with the numerical one, whereas the matching of the
homogeneous one is poorer. The optimal values of the variational
parameters $(\kappa_v,\kappa_\phi,\eta,h)$ for the piecewise choice
are $(2.57,2.6,0.98,0.15)$ for $\Xi=1$ and $(0.83,0.13,0.97,1.37)$
for $\Xi=100$.

\begin{figure}[t]
\includegraphics[width=1\linewidth]{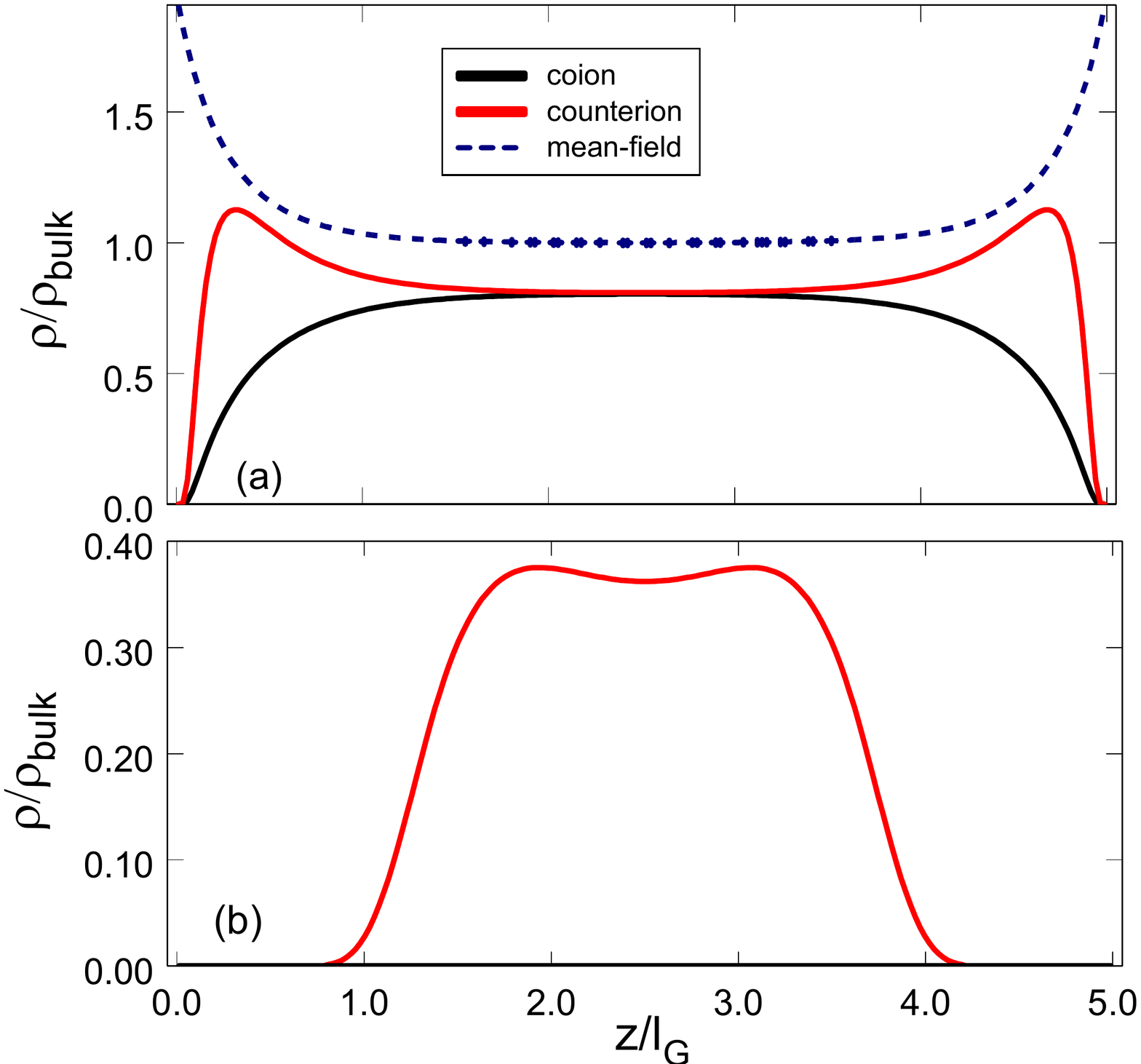}
\caption{(color online) Local ionic partition coefficient in the
nanopore (same parameters as in Fig.~\ref{FieldSL} and $\epsilon=0$)
computed with the piecewise solution. (a) $\Xi=1$ , (b) $\Xi=100$.
The dotted line in the top plot corresponds to the mean-field
prediction for counterion density.} \label{DENslab}
\end{figure}
The form of the electrostatic potential $\phi_0(z)$ is intimately
related to ionic concentrations. Ion densities inside the pore are
plotted in Fig.~\ref{DENslab} for $\Xi=1$ and $\Xi=100$. We first
notice that even at $\Xi=1$, the counterion density is quite
different from the mean-field prediction. Furthermore, due to image
charge and electrostatic repulsions from both sides, the coion
density has its maximum in the middle of the pore. On the other
hand, the counterion density exhibits a double peak, symmetric with
respect to the middle of the pore, which originates from the
attractive force created by the fixed charge and the repulsive image
forces. When $\Xi$ increases, we see that the counterion density
close to the wall shrinks and becomes practically flat in the middle
of the pore. Hence the potential $\phi_0$ linearly increases with
$z$ until the counterion-peak is reached and then it remains almost
constant since the counterion layer screens the electrostatic field
created by the surface charge (since in Fig.~\ref{FieldSL}, $z$ is
renormalized by the Gouy-Chapman length which decreases with
increasing $\sigma_s$, one does not see the increase of the slope
$\phi_0(z=0)$). In agreement with the variational Donnan
approximation above, coions are totally excluded from the pore for
large $\Xi$. Hence, the piecewise potential allows one to go beyond
the variational Donnan approximation within which the density
profile does not exhibit any concentration peak.

The inverse screening length $\kappa_v$ obtained with the piecewise
solution is compared in Fig.~\ref{KPvsKSI} with the prediction of
the Donnan approximation and that of the numerical solution. The
agreement between piecewise and numerical solutions is extremely
good. Although the Donnan approximation slightly underestimates the
salt density in the pore, its predictions follow the correct trend.

\section{Conclusion}

In this study, we applied the variational method to interacting
point-like ions in the presence of dielectric discontinuities and
charged boundaries. This approach interpolates between the WC limit
($\Xi\ll1$) and the SC one ($\Xi\gg1$), originally defined for
charged boundaries without dielectric discontinuity, and takes into
account image charge repulsion and solvation effects. The
variational Greens's function $v_0$ has a Debye-H\"uckel form with a
variational parameter $\kappa_v$ and the average variational
electrostatic potential $\phi_0(z)$ is either computed numerically
or a restricted form is chosen with variational parameters. The
physical content of our restricted variational choices can be
ascertained by inspecting the general variational equations
(Eqs.~(\ref{VarNetzP})-(\ref{VarNetzG}) for symmetric salts). The
generalized Onsager-Samaras approximation that we have adopted for
the Green's function replaces a local spatially varying screening
length by a constant variational one; although near a single
interface this screening length is equal to the bulk one, in
confined geometries the constant variational screening length can
account in an average way for the modified ionic environment (as
compared with the external  bulk with which the pore is in
equilibrium) and can therefore strongly deviate from the bulk value.
This modified ionic environment arises both from dielectric and
reduced solvation effects present even near neutral surfaces
(encoded in the Green's function) and the surface charge effects
encoded in the average electrostatic potential. Our restricted
variational choice for $\phi_0(z)$  is based on the usual non-linear Poisson-Boltzmann type
solutions with a renormalized inverse screening length that may
differ from the one used for $v_0$ and a renormalized external
charge source. The coupling between $v_0$ and $\phi_0$ arises
because the inverse screening length for $v_0$ depends on $\phi_0$
and vice-versa. The optimal choices are the ones that extremize the
variational free energy.

In the first part of the work, we considered single interface
systems. For asymmetric electrolytes at a single neutral interface,
the potential $\phi_0(z)$ created by charge separation was
numerically computed. It was satisfactorily compared to a restricted
piecewise variational solution and both charge densities and surface
tension are calculated in a simpler way than Bravina~\cite{bravina}
and valid over a larger bulk concentration range. The variational
approach was then applied to a single charged surface and it was
shown that a piecewise solution, characterized by two zones, can
accurately reproduce the correlations and non-linear effects
embodied in the more general variational equation. The first zone of
size $h$ is governed by a salt-free regime, while the second region
corresponds to an effective mean-field limit. The variational
calculation predicts a relation between $h$ and the surface charge
of the form $h\propto(c+\ln|\sigma_s|)/|\sigma_s|$ where the
parameter $c$ depends on the temperature and ion valency.

In the second part, we dealt with a symmetric electrolyte confined
between two dielectric interfaces and investigated the important
problem of ion rejection from neutral and charged membranes. We
illustrated the effects of ion valency and dielectric discontinuity
on the ion rejection mechanism by focusing on ion partition and salt
reflection coefficients. We computed within a variational Donnan
potential approximation, the inverse internal screening length and
ion partition coefficients, and showed that for $\Xi>4$ one reaches
the SC limit, where the partition coefficients are independent of
the bulk concentration and depend only on the size and charge of the
nanopore. This result has important experimental applications, since
it indicates that complete filtration can be done at low bulk salt
concentration and/or high surface charge. Furthermore, we showed
that, due to image interactions, the quantity of salt allowed to
penetrate inside a \textit{neutral} nanopore increases with the
pore-size. In the case of strongly charged membranes, this behavior
is reversed for the whole physical range of pore size. We quantified
the interplay between the image charge repulsion and the surface
charge attraction for counterions and found that even in the
presence of a weak surface charge, the competition between them
leads to a characteristic pore size $d_{cr}$ below which the
counterion partition coefficient rapidly decreases with increasing
pore size. On the other hand, for nanopores of size larger than
$d_{cr}$ the system behaves like a neutral pore. Our variational
calculation was compared to the Debye closure approach and the
midpoint approximation used by Yaroshchuk~\cite{yarosch}. The
closure equations have no exact solution even at the numerical
level. Our approach, based on restricted variational choices, shows
significant deviations from Yaroshchuk's mid-point approach at high
ion concentrations and small pore size. Finally, the introduction of
a simple piecewise trial potential for $\phi_0$, which perfectly
matches the numerical solutions of the variational equations,
enabled us to go beyond the variational Donnan potential
approximation and thus account for the concentration peaks in
counterion densities. We computed  ion densities in the pore and
showed that for $\Xi>4$, the exclusion of coions from the pore is
total. We also compared the ionic density profiles obtained from the
variational method with MC simulation results and showed that the
agreement is quite good, which illustrates the accuracy of the
variational approach in handling the correlation effects absent at
the mean-field level.

The main goal in this work was first to connect two different fields
in the chemical physics of ionic solutions focusing on complex
interactions with surfaces: field-theoretic calculations and
nanofiltration studies. Moreover, on the one hand, this variational
method allows one to consider, in a non-perturbative way,
correlations and non-linear effects; on the other hand the choice of
one constant variational Debye-H\"uckel parameter is simple enough
to reproduce previous results and to illuminate the mechanisms at
play. This approach is also able to handle, in a very near future,
more complicated geometries, such as cylindrical nanopores, or a
non-uniform surface charge distribution.

The present variational scheme also neglects ion-size effects and
gives rise to an instability of the free energy at extremely high
salt concentration. Second order corrections to the variational
method may be necessary in order to properly consider ionic
correlations leading to pairing~\cite{Fisher,Simonin} and to
describe the physics of charged liquids at high valency, high
concentrations or low temperatures. Introducing ion size will also
allow us to introduce an effective dielectric permittivity
$\epsilon_p$ for water confined in a nanopore intermediate between
that of the membrane matrix and bulk water, leading naturally to a
Born-self energy term that varies inversely with ion size and
depends on the difference between $1/\epsilon_w$ and
$1/\epsilon_p$~\cite{Senap,Marti}. Furthermore, the incorporation of
the ion polarizability \cite{Dipoles} will yield a more complete
physical description of the behavior of large ions~\cite{Jungwirth}.
Charge inversion phenomena for planar and curved interfaces is
another important phenomenon that we would like to consider in the
future~\cite{Lorenz}. Note, however, that our study of asymmetric
salts near neutral surfaces reveals a closely related phenomenon:
the generation of an effective non-zero surface charge due to the
unequal ionic response to a neutral dielectric interface for
asymmetric salts. A further point that possesses experimental
relevance is the role played by surface charge inhomogeneity. Strong-coupling
calculations show that an inhomogeneous surface charge distribution
characterized by a vanishing average value gives rise to an
attraction of ions towards the pore walls, but this effect
disappears at the mean-field level~\cite{NajiDIS}. For a better
understanding of the limitations of the proposed model, more
detailed comparison with MC/MD simulations are in
order\cite{Ge,Leung}. Finally, dynamical hindered transport
effects~\cite{lefebvreII,Ge} such as hydrodynamic forces deserve to
be properly included in the theory for practical applications.
\\
\acknowledgements We would like to thank David S. Dean for numerous
helpful discussions. This work was supported in part by the French
ANR Program NANO-2007 (SIMONANOMEM project, ANR-07-NANO-055).

\smallskip

\appendix
\section{Variational free energy}
\label{appendixVarFr}

For planar geometries (charged planes), the translational invariance
parallel to the plane, allows us to significantly simplify the
problem by introducing the partial Fourier-transformation of the
trial potential in the form \be\label{FourTr}
v_0(z,z',\br_{||}-\br'_{||})=\int\frac{\mathrm{d}\bk}{(2\pi)^2}e^{i\bk\cdot\left(\br_{||}-\br'_{||}\right)}
\hat{v}_0(z,z',\bk). \ee By injecting the Fourier decomposition
(\ref{FourTr}) into Eq.~(\ref{DH1}), the DH equation becomes \bea
&&\left\{-\frac{\partial}{\partial z}\epsilon(z)\frac{\partial}{\partial z}+\epsilon(z)[k^2+\kappa^2_v(z)]\right\}\hat{v}_0\left(z,z',\bk;\kappa_v(z)\right)\nonumber\\
&&\hspace{4cm}=\frac{e^2}{k_BT}\delta(z-z'). \label{DHFour} \eea

The translational symmetry of the system enables us to express any
thermodynamic quantity in terms of the partially Fourier-transformed
Green's function $\hat{v}_0(z,z,\bk)$. The average electrostatic
potential contribution to $F_v$ that follows from the average
$\left<H\right>_0$ reads \bea\label{F12}
F_1&=&S\int \mathrm{d}z\left\{-\frac{\left[\nabla\phi_0(z)\right]^2}{8\pi \ell_B}+\rho_s(z)\phi_0(z)\right.\nonumber\\
&&-\left.\sum_i\lambda_i
e^{-\frac{q_i^2}{2}W(z)-q_i\phi_0(z)}\right\}, \eea the kernel part
is \bea\label{FvarIV}
F_2&=&\frac{S}{16\pi^2}\int_0^1 \mathrm{d}\xi\int_0^\infty \mathrm{d}k k\int \mathrm{d}z\frac{\kappa^2_v(z)}{\ell_B(z)}\\
&\times &
\left[\hat{v}_0\left(z,z,\bk;\kappa_v(z)\sqrt{\xi}\right)-\hat{v}_0\left(z,z,\bk;\kappa_v(z)\right)\right]\nonumber
\eea where the first term in the integral follows from $F_0$ and the
second term from $\left<H_0\right>_0$.  Finally, the unscreened Van
der Waals contribution, which comes from the unscreened part of
$F_0$ , is given by \bea\label{Fc}
F_3&=&\frac{S}{8\pi}\int_0^1 \mathrm{d}\xi\int_0^\infty \mathrm{d}z\left[\frac{1}{\ell_B(z)}-\frac{1}{\ell_B}\right]\left<\left(\nabla\phi\right)^2\right>_\xi\nonumber\\
&&-\ln\int\mathcal{D}\phi\hspace{0.5mm}e^{-\int\frac{\mathrm{d}\br}{8\pi\ell_B}\left(\nabla\phi\right)^2}
\eea The technical details of the computation of $F_3$ can be found
in Ref.~\cite{netz_vdw}. The last term of Eq. (\ref{Fc}) simply
corresponds to the free energy of a bulk electrolyte with a
dielectric constant $\epsilon_w$. In the above relations, $S$ stands
for the lateral area of the system. The dummy ``charging'' parameter
$\xi$ is usually introduced to compute the Debye-H\"uckel free
energy~\cite{mcquarrie}. It multiplies the Debye lengths of
$\hat{v}_0\left(z,z,\bk;\kappa_v(z)\right)$ in Eq.~(\ref{FvarIV})
and the dielectric permittivities contained in the thermal average
of the gradient in Eq.(\ref{Fc}). This later is defined as \bea
\left\langle\left(\nabla\phi\right)^2\right\rangle_\xi &=& -\left(\nabla\phi_0\right)^2 \label{AvC}\\
&+&\int\frac{\mathrm{d}\bk}{(2\pi)^2}\left(k^2+\partial_z\partial_{z'}\right)
\left.\hat{v}_c\left[z,z',\bk;\ell_\xi(z)\right]\right|_{z=z'}\nonumber
\eea where we have introduced
$\ell_\xi^{-1}(z)\equiv\ell_B^{-1}+\xi\left[\ell_B^{-1}(z)-\ell_B^{-1}\right]$
and $\hat{v}_c\left[z,z',\bk;\ell_\xi(z)\right]$ stands for the
Fourier transformed  Coulomb operator given by Eq.~(\ref{coulomb})
with Bjerrum length $\ell_\xi(z)$. The quantity $F_3$ defined in
Eq.~(\ref{Fc}) does not depend on the inverse screening length
$\kappa_v$. Moreover, in order to satisfy the electroneutrality,
$\phi_0(z)$ must be constant in the salt-free parts of the system
where $\ell_B(z)\neq \ell_B$. Hence, $F_3$ does not depend on the
potential $\phi_0(z)$.

\section{Variational choice for the neutral dielectric interface}
\label{appendixWA}

We report in this appendix the restricted variational piecewise
$\phi_0(z)$ for a neutral dielectric interface which is a solution
of \bea
\frac{\partial^2 \phi_0}{\partial z^2}&=&0\quad\mathrm{for}\quad z\leq a\label{eqvarIV1}\\
\frac{\partial^2 \phi_0}{\partial
z^2}&-&\kappa_\phi^2\phi_0=cze^{-\kappa_\phi
z}\quad\mathrm{for}\quad z\geq a \label{eqvarIV2} \eea where
$\phi_0(z)$ in both regions is joined by the continuity conditions
\be
\phi^<_0(a)=\phi^>_0(a),\hspace{1cm}\left.\frac{\partial\phi^<_0}{\partial
z}\right|_{z=a}=\left.\frac{\partial\phi^>_0}{\partial
z}\right|_{z=a}.\label{Cont} \ee We also tried to introduce
different variational screening lengths in the second term of the
lhs. and in the rhs. of Eq.~(\ref{eqvarIV2}) without any significant
improvement at the variational level. For this reason, we opted for
a single inverse variational screening length, $\kappa_\phi$. The
solution of Eqs.~(\ref{eqvarIV1})-(\ref{eqvarIV2}) is \be\label{TriPotWA2}
\phi_0(z)=\left\lbrace
\begin{array}{ll}
\varphi & \mathrm{for}\quad z\leq a,\\
\varphi\left[1+\kappa_\phi (z-a)\right]e^{-\kappa_\phi(z-a)} &
\mathrm{for}\quad z\geq a.
\end{array}\right.
\ee where the coefficient $c$ disappears when we impose the boundary
and continuity conditions, Eq.~(\ref{Cont}). The remaining
variational parameters are the constant potential $\varphi$, the
distance $a$ and the inverse screening length $\kappa_\phi$. By
substituting Eq.~(\ref{TriPotWA2}) into Eq.~(\ref{F12}), we obtain
the variational grand potential \bea\label{FreeWA}
&&F_v=V\frac{\kappa_b^3}{24\pi}+\frac{S}
{32\pi}\left(\Delta\kappa_b^2-\frac{\kappa_\phi}{\ell_B}\varphi^2\right)
-S\rho_-\int_0^\infty \mathrm{d}z \nonumber \\ &&\times
\left\{e^{-\frac{q_-^2}{2}w(z)+q_-\phi_0(z)}+\frac{q_-}{q_+}e^{-\frac{q_+^2}{2}w(z)-q_+\phi_0(z)}\right\}
\eea

\section{Variational choice for the charged dielectric interface}
\label{appendixIC}

The two types of piecewise variational functions used for single
charged surfaces are reported below.
\begin{itemize}
   \item The first trial potential obeys the salt-free equation in the first zone and the NLPB solution in the second zone,
\bea\label{eqvarVI}
\frac{\partial^2 \tilde\phi_0^{\rm NL}}{\partial\tilde{z}^2}&=&2\delta(\tilde{z})\quad\mathrm{for}\quad\tilde{z}\leq\tilde{h},\\
\frac{\partial^2 \tilde\phi_0^{\rm NL}}{\partial
\tilde{z}^2}&-&\tilde{\kappa}_\phi^2\sinh\phi_0=0\quad\mathrm{for}\quad
\tilde{z}\geq \tilde{h},\nonumber \eea whose solution is
\be\label{resNL2}
\tilde \phi_0^{\rm NL}(\tilde{z})=\left\lbrace
 \begin{array}{ll}
 4\mathrm{arctanh}\gamma+2(\tilde{z}-\tilde{h})&\mathrm{for}\quad\tilde{z}\leq \tilde{h},\\
 4\mathrm{arctanh}\left(\gamma e^{-\tilde{\kappa}_\phi(\tilde{z}-\tilde{h})}\right)&\mathrm{for}\quad \tilde{z}\geq \tilde{h},
 \end{array}\right.
\ee where
$\gamma=\tilde{\kappa}_\phi-\sqrt{1+\tilde{\kappa}_\phi^2}$.
Variational parameters are $h$ and $\kappa_\phi$, and the
electrostatic contribution of the variational grand potential
Eq.~(\ref{F12}) is \be
\frac{F_1}{\tilde{S}}=\frac{\tilde{h}+\gamma-4\mathrm{arctanh}\gamma}{2\pi\hspace{0.5mm}\Xi}-
\frac{\tilde{\kappa}_b^2}{4\pi\Xi}\int d\tilde{z}
e^{-\frac{\Xi}{2}\tilde w(\tilde{z})}\cosh\tilde\phi_0^{\rm NL}. \ee

\item The second type of trial potential obeys the salt-free equation with a charge renormalization in the first zone and the linearized Poisson-Boltzmann solution in the second zone,
\bea\label{eqvarVII}
\frac{\partial^2 \tilde\phi_0^{\rm L}}{\partial \tilde{z}^2}&=&2\eta\delta(\tilde{z})\quad\mathrm{for}\quad\tilde{z}\leq \tilde{h},\nonumber\\
\frac{\partial^2 \tilde\phi_0^{\rm L}}{\partial
\tilde{z}^2}&-&\tilde{\kappa}_\phi^2\tilde\phi_0^{\rm
L}=0\quad\mathrm{for}\quad\tilde{z}\geq \tilde{h}, \eea whose
solution is given by \be
\tilde \phi_0^{\rm L}(\tilde{z})=\left\lbrace
 \begin{array}{ll}
 -\frac{2\eta}{\tilde{\kappa}_\phi}+2\eta(\tilde{z}-\tilde{h})& \mathrm{for}\quad\tilde{z}\leq\tilde{h},\\
 -\frac{2\eta}{\tilde{\kappa}_\phi}e^{-\tilde{\kappa}_\phi(\tilde{z}-\tilde{h})}&\mathrm{for}\quad \tilde{z}\geq\tilde{h}.
 \end{array}\right.
 \label{resL2}
\ee Variational parameters introduced in this case are $\tilde h$,
$\tilde\kappa_\phi$, and the charge renormalization $\eta$, which
takes into account non-linearities at the mean-field
level~\cite{netz_var}. The variational grand potential reads \bea
\frac{F_1}{\tilde{S}}&=&\frac{2\eta(1+\tilde{h}\tilde{\kappa}_\phi)-\eta^2(1/2+\tilde{h}\tilde{\kappa}_\phi)}{2\pi\Xi\tilde{\kappa}_\phi}
\nonumber\\
&-&\frac{\tilde{\kappa}_\phi^2}{4\pi\Xi}\int d\tilde{z}
e^{-\frac{\Xi}{2}\bar w(\tilde{z})}\cosh\tilde\phi_0^{\rm L}(\tilde
z). \eea
\end{itemize}

In both cases, the boundary condition satisfied by $\phi_0$ is the
Gauss law \be \left.\frac{\partial\tilde\phi_0}{\partial \tilde
z}\right|_{z=0}=2\eta \label{Gauss} \ee where $\eta=1$ for the
non-linear case. It is important to stress that in the case of a
charged interface, Eq.~(\ref{Gauss}) holds even if $\epsilon\neq 0$.
In fact, since the left half-space is ion-free, $\tilde\phi_0(z)$
must be constant for $z<0$ in order to satisfy the global
electroneutrality in the system.

\section{Definition of the special functions}
\label{appendix}

The definition of the four special functions used in this work are
reported below. \bea
{\rm Li}_n(x) &=& \sum_{k\geq1}\frac{x^k}{k^n},\quad \xi(n)={\rm Li}_n(1)\\
\beta(x;y,z) &=& \int_0^x \mathrm{d}t\; t^{y-1}(1-t)^{z-1}\\
_2\mathrm{F}_1(a,b;c;x) &=&
\sum_{k\geq0}(a)_k(b)_k(c)_k\frac{x^k}{k!} \eea where
$(a)_k=a!/(a-k)!$.

\section{Disjoining pressure for the neutral pore}
\label{appendixPR}
\begin{figure}
\includegraphics[width=1\linewidth]{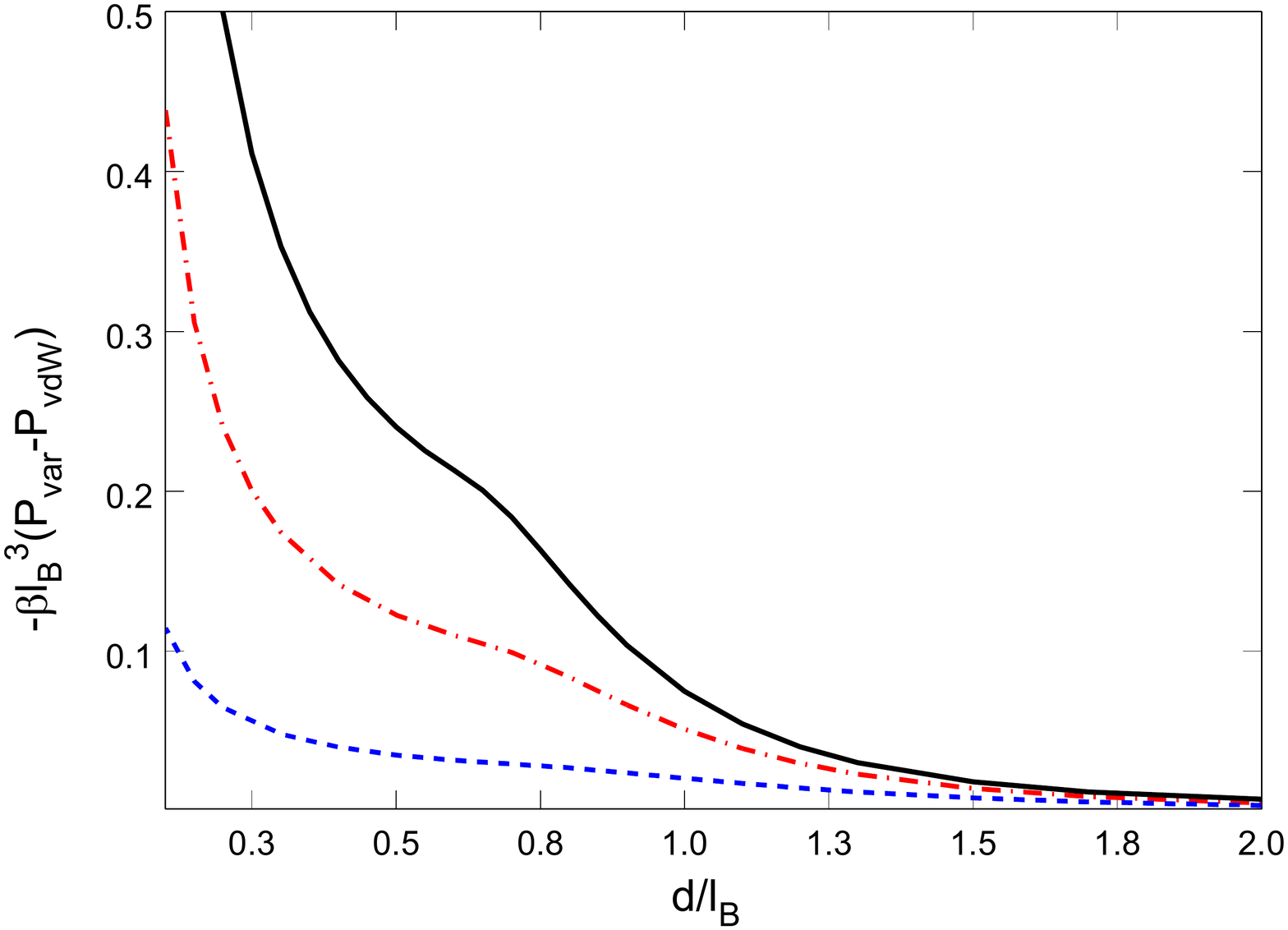}
\caption{(color online) Difference between the pressure and the
screened van der Waals contribution vs $d/\ell_B$ for
$\kappa_bl_B=0.5,1$  and 1.5, from left to right ($\epsilon=0$).}
\label{PressureN}
\end{figure}
The net pressure between plates is defined as \be\label{PrI}
P=-\frac1S\frac{\partial F_v}{\partial
d}-\left(2\rho_b-\frac{\kappa_b^3}{24\pi}\right) \ee where the
subtracted term on the rhs. is the pressure of the bulk electrolyte.
The total van der Waals free energy, which is simply the zeroth
order contribution $F_0$ to the variational grand potential
Eq.~(\ref{Fvar}), is with the constraint $\kappa_v=\kappa_b$ (there
is no renormalization of the inverse screening length at this
order), \be\label{vdW} F_{\rm vdW}=\frac{d\kappa_b^3}{24\pi}
-\frac{\kappa_b}{8\pi d}\mathrm{Li}_2\left(e^{-2d\kappa_b}\right)
-\frac1{16\pi d^2}\mathrm{Li}_3\left(e^{-2d\kappa_b}\right) \ee and
\be P_{\rm vdW}=-\frac{1}{S}\frac{\partial F_{\rm vdW}}{\partial
d}+\frac{\kappa_b^3}{24\pi}. \ee We illustrate in
Fig.~\ref{PressureN} the difference between the van der Waals
pressure and the prediction of the variational calculation for
$\kappa_b\ell_B=0.5$, 1 and 1.5. We notice that the prediction of
our variational calculation yields a very similar behavior to that
illustrated in Fig.~8 of Ref.~\cite{hatlo}. The origin of the
extra-attraction that follows from the variational calculation was
discussed in detail in the same article. This effect originates from
the important ionic exclusion between the plates at small interplate
separation, an effect that can be captured within the variational
approach.


\begin{thebibliography}{99}
\bibitem{footnote1} Note that the factor $\Xi^{-1}$ in front of $v_0$ in Eq.~(\ref{VarNetzG}) found in~\cite{netz_var} should be cancelled.
\end{thebibliography}
\end{document}